\documentclass{IEEEtran}

\usepackage{amsmath,amsthm,amsfonts,amscd,amssymb, arydshln}
\usepackage[mathscr]{eucal}
\usepackage{graphics,graphicx,multicol}
\usepackage{epsfig}
\usepackage{diagbox}
\usepackage{subfigure}
\usepackage{stfloats}
\usepackage{latexsym}
\usepackage{amsfonts}
\usepackage{cite}
\usepackage{algorithm,algorithmic}
\usepackage{color}
\usepackage{xcolor}
\usepackage{multirow}
\usepackage{booktabs}

\usepackage[table]{xcolor}

\usepackage{nicematrix}
\usepackage{arydshln} 

\usepackage{booktabs}
\usepackage{array}
\usepackage{tabularray}
\usepackage{arydshln}
\usepackage{arydshln,leftidx,mathtools}

\usepackage{amsmath,amssymb,amsfonts}
\usepackage{algorithmic}
\usepackage{graphicx}
\usepackage{textcomp}
\usepackage{xcolor}
\usepackage{amsmath}
\usepackage{amssymb}
\usepackage{mathtools}
\usepackage{amsthm}
\usepackage{amsfonts}
\usepackage{stackengine}

\newtheorem{theorem}{Theorem}

\newtheorem{corollary}{Corollary}
\newtheorem{lemma}{Lemma}

\usepackage{float}

\DeclarePairedDelimiterX{\inp}[2]{\langle}{\rangle}{#1, #2}

\IEEEoverridecommandlockouts

\begin{document}

\title{Multi-Antenna Users in Cell-Free Massive MIMO: Stream Allocation and Necessity of Downlink Pilots
\thanks{This work was supported in part by the SUCCESS project funded by the Swedish Foundation for Strategic Research; in part by the Swedish Foundation for Strategic Research (SSF) under Grant IS24-0190; in part by the Basic Research Program-Individual Research (Global Matching Grants-Sweden) through the National Research Foundation (NRF) of Korea, funded by the Korean government (Ministry of Science and ICT) under Grant RS-2024-00436923. A preliminary version of this work was presented at the IEEE Wireless Communications and Networking Conference,
21–24 April 2024. \textcolor{black}{\textit{(Corresponding Authors: Eren Berk Kama; Junbeom Kim.)}}}
}

 \author{Eren Berk Kama,~\IEEEmembership{Student Member,~IEEE}, Junbeom Kim,~\IEEEmembership{Member,~IEEE}, and Emil Bj{\"o}rnson,~\IEEEmembership{Fellow,~IEEE}
\thanks{Eren Berk Kama and Emil Bj{\"o}rnson are with the Division of Communication Systems, KTH Royal Institute of Technology, Stockholm, Sweden (e-mail: ebkama@kth.se; emilbjo@kth.se).

Junbeom Kim is with the Department of AI Information Engineering, Gyeongsang National University, Jinju, South Korea (e-mail: junbeom@gnu.ac.kr).}}
\maketitle
\date{}
\begin{abstract}
\textcolor{black}{We consider a cell-free massive multiple-input multiple-output (MIMO) system with multiple antennas on the users and access points (APs). In previous works, the downlink spectral efficiency (SE) has been evaluated using the hardening bound that requires no downlink pilots. This approach works well for single-antenna users. In this paper, we show that much higher SEs can be achieved if downlink pilots are sent when having multi-antenna users. The reason is that the effective channel matrix does not harden. We propose a pilot-based downlink estimation scheme, derive a new SE expression, and show numerically that it yields substantially higher performance when having correlated Rayleigh fading channels. }

\textcolor{black}{
In cases with multi-antenna users, the APs can either coherently transmit the same data streams, or alternatively they transmit separate data streams non-coherently. 
The latter approach reduces the fronthaul signaling overhead, albeit with a potential SE penalty. For both strategies, we propose novel precoding and combining schemes.
Specifically, we develop precoders based on the minimum mean square error (MMSE) criterion and investigate receive combining methods, including an MMSE combiner. Furthermore, we consider sharing different levels of channel knowledge between the APs.
Finally, we present a comprehensive numerical analysis to validate our findings, evaluating the performance trade-offs associated with the number of users, APs, and antennas, as well as the choice of transmission strategy and the level of CSI sharing among APs. 
}



\end{abstract}


\section{Introduction}

Cell-free massive multiple-input multiple-output (MIMO) is a wireless communication paradigm that has attracted great interest due to the vision of delivering uniformly high spectral efficiency (SE) over the coverage area \cite{interdonato2019ubiquitous,chen2022survey,9650567}.
In cell-free massive MIMO, a large number of geographically distributed access points (APs) cooperate to serve the users on the same time-frequency resources through coherent joint transmission where interference is managed by MIMO methods \cite{demir2021foundations}.

The recent surveys \cite{9650567,chen2022survey} highlight a large body of work on cell-free massive MIMO systems. The predominant assumption is single-antenna users, although user devices (e.g., phones and tablets) have been equipped with multiple antennas since 4G. Four antennas are the current norm for mid-band 5G devices, and larger antenna numbers are expected in 6G devices operating at higher frequencies.
The few papers on cell-free massive MIMO that consider multi-antenna users show SE improvements  (e.g., \cite{buzzi2017cell}) but assume perfect channel state information (CSI).
A detailed SE analysis with realistic imperfect CSI is missing, particularly for downlink operation.

The uplink SE with zero-forcing (ZF) combining and multi-antenna users was studied in \cite{mai2019uplink} under the assumption of independent and identically distributed (i.i.d.) Rayleigh fading. This analysis was extended to consider the Weichselberger channel model in \cite{wang2022uplink}, where four operation regimes were compared. The uplink was further studied in \cite{zhang2019performance,sun2023uplink} with a focus on distortion caused by low-resolution hardware.
The downlink performance with ZF precoding is analyzed in \cite{buzzi2019user} under i.i.d. Rayleigh fading, using the hardening bound where the user device lacks CSI. \textcolor{black}{Moreover, the downlink performance with maximum ratio (MR) precoder and downlink channel estimation was studied in \cite{mai2020downlink}. Furthermore, closed-form SE relations are given for the particular assumptions in their setting.}\textcolor{black}{While it is a foundational analysis of downlink SE in cell-free massive MIMO with multi-antenna users, its reliance on the assumption that the effective channel matrices converge to Gaussian matrices prevents generalization beyond simple precoders like the MR precoder. }
The downlink with low-resolution digital-to-analog converters (DACs) was considered in \cite{zhou2021sum}.
Multi-antenna users have also been studied in the context of cellular massive MIMO.
A single-cell scenario was considered in \cite{li2016massive} and the hardening bound was used to compute an achievable downlink SE without CSI at the users. 
\cite{dovelos2020massive} computed a closed-form uplink SE expression in single-cell massive MIMO systems with multi-antenna users, MR combining, and Rician fading.
Downlink MR precoding with multi-antenna users was considered in \cite{sutton2021hardening}, with a focus on making the hardening bound tighter with an unusual precoding normalization.
\textcolor{black}{\cite{10571171} investigates the impact of spatial correlation and mutual coupling on SE.} \textcolor{black}{
Whereas \cite{wang2020uplink} analyzes uplink SE over spatially correlated Rician fading channels, comparing maximum ratio (MR) and local MMSE combining schemes. 
} The correlation between the channel vectors to different antennas at the same user was studied in \cite{wu2017novel}, and it was shown that asymptotic orthogonality cannot be guaranteed.
Synchronization issues, such as phase misalignments between APs, were studied in \cite{ganesan2024cell} by proposing transmission methods and optimization of the AP clusters.

In \cite{ngo2017no}, it was shown that no downlink pilots are needed in massive MIMO systems with single-antenna users. The receiving users can perform decoding properly without CSI since the effective channel's phase is removed by precoding, and the gain approaches its mean value---a phenomenon called channel hardening. The so-called hardening bound on the capacity gives a performance close to the perfect CSI  case also in cell-free massive MIMO \cite{demir2021foundations} with the possible exception of scenarios with MR precoding, if the precoding vectors are normalized incorrectly \cite{Chen2018a,Interdonato2021a} and when the effective channel means are normalized with their estimates \cite{8799031}. In cases with very limited channel hardening (e.g., keyhole channels), higher SE can be achieved by explicitly estimating the effective channel's amplitude from the received downlink signals \cite{ngo2017no}.
These single-antenna results might be the reason why the hardening bound has also been used in \cite{buzzi2019user,zhou2021sum,li2016massive,sutton2021hardening} when considering multi-antenna users. However, we will show that the use of downlink pilots can lead to great SE improvements in these cases. \textcolor{black}{ \cite{kanno2022fronthaul} proposes hybrid signal processing schemes for uplink cell-free massive MIMO, compressing received signals at APs to reduce fronthaul loads, while \cite{shaik2025over} introduces an over-the-air (OTA) computation framework to minimize fronthaul overhead by computing global sufficient statistics locally at APs. Both these solutions are post-processing methods, whereas our work introduces a fundamentally different precoding method.
}


\subsection{Motivation and Contributions}

In this paper, we analyze the performance of a cell-free massive MIMO system with multiple antennas on the users and APs, in a scenario with correlated Rayleigh fading channels.
Conventional cell-free massive MIMO builds on the assumption that the serving APs send the same data stream to the user. However, we can send multiple streams to multi-antenna users, which enables an alternative transmission method where the serving APs send different streams to the same user \textcolor{black}{non-coherently}. \textcolor{black}{This separate-stream transmission mode allows each AP to forward an independent data stream to the UE, thereby eliminating the need for network-wide phase calibration and slashing the downlink fronthaul load. It represents an operating point that is fundamentally impossible with single-antenna UEs and has not been explored in earlier cell-free research.} 
We analyze both transmission methods with arbitrary precoding to derive general SE expressions.
In particular, we derive SE expressions for the cases when the user has CSI obtained through pilot-based downlink channel estimation. \textcolor{black}{Unlike the scalar channels of single-antenna users, the effective MIMO channel seen by a multi-antenna UE is a matrix that does not harden to a deterministic scaled identity. Consequently, estimating this matrix via downlink pilots is indispensable, and our analysis shows that doing so closes the large spectral-efficiency gap between the hardening bound and the perfect-CSI benchmark.}

We propose new precoders for the two transmission methods. These are needed because the conventional precoding schemes for multi-antenna users in \cite{bandemer2006linear,stankovic2004multi,spencer2004zero}
are not designed for situations with per-AP power constraints, partial CSI, and separate stream transmission.
We further investigate different combining methods that suppress interference using the downlink channel estimates.



We provide numerical results that corroborate our results, including the importance of downlink pilots, and evaluate how different system parameters affect the SEs.

The main contributions of this paper are as follows:

\begin{itemize}

\item We introduce two downlink transmission methods for multi-antenna users: same and separate stream transmission.
We divide the latter method into two sub-methods depending on the CSI-sharing among the APs. We show under what conditions the same and separate stream transmission methods perform equally well. We propose precoders suitable for each case by following a minimum mean square error (MMSE) approach.

\item We provide a novel downlink SE expression achievable by using downlink pilots and explain why it gives much higher values than the hardening bound. We investigate the performance differences between the ZF and MMSE combiners. 

\item We demonstrate the performance gap between the hardening bound and the perfect CSI case numerically and show that the proposed SE expression with downlink channel estimation bridges the gap between them. 

\end{itemize}

\textcolor{black}{In the conference version of this work \cite{kama2024downlinkpilotsessentialcellfree}, we considered a substantially simpler setup with i.i.d. Rayleigh fading, centralized operation, and ZF combining, which was sufficient to demonstrate the necessity of downlink pilots when having multi-antenna users. By contrast, in this paper, we modify the channel model to correlated Rayleigh fading (which requires more advanced estimation and more general SE expressions), explore three different transmission methods (two of which are novel contributions), and multiple receive combining methods (both MMSE and ZF combining). Moreover, we propose precoders tailored to the different transmission methods and derive novel SE relations for the general combiner case. These extensions and generalizations make the contributions applicable in practical conditions and provide insight into which design choices to make.}

The rest of this paper is organized as follows. In Section \ref{System Model}, we describe the channel model and estimation of the channel in the uplink. In Section \ref{sec: Downlink Data Transmission}, downlink data transmission methods are explained for both same and separate stream methods. In Section \ref{sec: SE Results for Option 1}, achievable SE results for the same stream transmission method are derived, particularly with downlink pilots. In Section \ref{sec: SE Results for Option 2}, SE results are extended to the case with separate stream transmission. Next, in Section \ref{sec: Channel and Precoder Arguments for Option 2  Option 3}, precoders are proposed for all the considered transmission methods. We explain under which channel conditions these methods coincide. Section \ref{Numerical_Results} provides numerical results where we compare the performance of the considered methods and compare different SE expressions. Finally, conclusions and future directions are given in Section \ref{sec: Conclusions}.

\textit{Notation:} The superscripts $(\cdot)^{*}$,$(\cdot)^{\intercal}$ and $(\cdot)^{\mathrm{H}}$ denote the conjugate, transpose and conjugate transpose, respectively.  Column vectors and matrices are denoted with boldface lowercase $\mathbf{x}$ and uppercase $\mathbf{X}$ letters. The expectation operation, variance, and covariance of a vector and matrix are denoted as $\mathbb{E}\{\cdot\}, \mathrm{var}(\cdot),\mathrm{cov}(\cdot)$. \textcolor{black}{The covariances are evaluated as $\mathbf{C}_{\mathbf{a}\mathbf{b}}= \mathbb{E}\{\mathbf{a}\mathbf{b}^{\mathrm{H}}\}$ for $\mathbf{a}$ and $\mathbf{b}$ vectors.} The trace and determinant of a matrix are given by $\mathrm{tr}(\cdot), |\cdot|$. Convergence is shown with $\rightarrow$. The $L_{2}$ norm of a vector is given by $\Vert \mathbf{x} \Vert_{2}$ and the Frobenius norm of a matrix is given by $\Vert \mathbf{X} \Vert_{F}$. The $N$-dimensional identity matrix is denoted by $\mathbf{I}_{N}$. $\mathbf{x} \sim \mathcal{N}_{\mathbb{C}}(\mathbf{0},\mathbf{R})$ denotes a circularly symmetric complex Gaussian random vector $\mathbf{x}$ with covariance matrix $\mathbf{R}$.
The vectorization of a matrix $\mathbf{X}$ is denoted by $\mathrm{vec}(\mathbf{X})$.

\section{System Model} \label{System Model}
We consider a cell-free massive MIMO system with $L$ APs and $K$ users arbitrarily distributed in a large geographical area. Each AP has $N$ antennas and each user has $M$ antennas. We consider the standard block-fading time-division duplexing (TDD) operation where the channel between AP $l$ and user $k$ in an arbitrary coherence block is denoted by $\mathbf{H}_{lk} \in \mathbb{C}^{N \times M}$ \cite{demir2021foundations}. We assume a correlated Rayleigh fading channel model.  
Hence, the channel between AP $l$ and user $k$ is complex Gaussian distributed with the spatial correlation matrix $\mathbf{R}_{lk} \in \mathbb{C}^{NM \times NM}$, which is defined for the vectorization $\mathrm{vec}(\mathbf{H}_{lk})$ of the channel matrix $\mathbf{H}_{lk}$.  \textcolor{black}{
The channel covariance is expressed through the covariances of its individual column vectors. To obtain this covariance, we vectorize the channel matrix.}



\subsection{Uplink Channel Estimation}

As customary in TDD operation, the APs estimate the channels in the uplink and use the estimates in both uplink data reception and downlink data transmission \cite{demir2021foundations}. In this paper, we will focus on uplink channel estimation and downlink data transmission.
All users send uplink pilot sequences, and the APs estimate their channels to the users based on their received signals. Each user sends a $\tau_{\mathrm{p}}$-length orthonormal pilot sequence per transmit antenna. We will let $\tau_{\mathrm{c}}$  denote the length of the coherence block (in the number of symbols); thus, the pilot signaling overhead is $\tau_{\mathrm{p}}/\tau_{\mathrm{c}}$.

We denote the pilot matrix used by user $k$ as $\boldsymbol{\Phi}_{k} \in \mathbb{C}^{\tau_{\mathrm{p}} \times M}$. Since the coherence blocks are finite-sized, we assume that more than one user might be assigned the same pilot matrix. We use the set $\mathcal{P}_{k}$ to denote the set of users that share a pilot matrix with user $k$. The set of pilot matrices is selected so that $\boldsymbol{\Phi}_{i}^{\rm{H}}\boldsymbol{\Phi}_{i'}= \mathbf{I}_{M}$ if $i , i'\in \mathcal{P}_{k}$ and $\mathbf{0}$ otherwise, and user $k$ transmits $\sqrt{\tau_{\mathrm{p}}} \boldsymbol{\Phi}_{k}$ to make the pilot energy proportional to the pilot length.  The received signal at AP $l$ is 
\begin{equation}
    \mathbf{Y}_{l}^{\text{Pilot}}=\sum_{i=1}^{K} \sqrt{ q_{i} \tau_{\mathrm{p}}} \mathbf{H}_{li} \boldsymbol{\boldsymbol{\Phi}}_{i}^{\rm{H}} + \mathbf{N}_{l},
\end{equation}
where $q_{i}$ is the transmit power used by user $i$ normalized by the noise power and $\mathbf{N}_{l}$ is the $\textcolor{black}{N} \times \tau_{\mathrm{p}}$ noise matrix at AP $l$ with i.i.d. $\mathcal{N}_{\mathbb{C}}(0,1)$-entries. After correlating the received signal with the pilot matrix of user $k$, we obtain
\begin{align} \nonumber
    \mathbf{Y}_{lk} &=\mathbf{Y}_{l}^{\text{Pilot}} \boldsymbol{\Phi}_{k} \\ \nonumber
    &= \sqrt{q_k \tau_{\mathrm{p}} } \mathbf{H}_{lk} + 
    \sum_{i=1, i\neq k}^{K} \sqrt{q_i  \tau_{\mathrm{p}}} \mathbf{H}_{li} \left( \boldsymbol{\Phi}_{i}^{\rm{H}}\boldsymbol{\Phi}_{k}\right) + 
    \mathbf{N}_{lk} \\
     &= \sqrt{q_k \tau_{\mathrm{p}}} \mathbf{H}_{lk} + \sum_{{i \in \mathcal{P}_k \setminus \{ k\}}} \sqrt{q_i \tau_{\mathrm{p}}} \mathbf{H}_{li} + \mathbf{N}_{lk}, \label{eq:Y_lk}
\end{align}
where the second equality follows since the pilots used by user $k$ and user $i$ are orthogonal if $i \not \in \mathcal{P}_{k}$. This is a sufficient statistic for estimating $\mathbf{H}_{lk}$.
We notice that the new noise matrix $\mathbf{N}_{lk}=\mathbf{N}_{l}\boldsymbol{\Phi}_{k}$ also has i.i.d. $\mathcal{N}_{\mathbb{C}}(0,1)$-entries since $\boldsymbol{\Phi}_{k}$ is unitary; thus, every entry of $\mathbf{Y}_{lk}$ contains an observation of the corresponding Gaussian distributed entry of $\mathbf{H}_{lk}$ plus independent pilot interference and noise. We vectorize the result in \eqref{eq:Y_lk} and obtain
\begin{align}
    \mathbf{y}_{lk} &= \mathrm{vec}(\mathbf{Y}_{lk}) \nonumber \\ &= \sqrt{q_k \tau_{\mathrm{p}}} \mathrm{vec}(\mathbf{H}_{lk}) + \hspace{-0.5em}
    \sum_{{i \in \mathcal{P}_k \setminus \{ k\}}} \hspace{-0.5em} \sqrt{q_i \tau_{\mathrm{p}}} \mathrm{vec}(\mathbf{H}_{li})  + \mathrm{vec}(\mathbf{N}_{lk}) \nonumber
    \\
     &= \sqrt{q_k \tau_{\mathrm{p}}} \mathbf{h}_{lk} + \sum_{{i \in \mathcal{P}_k \setminus \{ k\}}} \sqrt{q_i \tau_{\mathrm{p}}} \mathbf{h}_{li} + \mathbf{n}_{lk},
\end{align}
where we introduced the notation $\mathbf{h}_{li} = \mathrm{vec}(\mathbf{H}_{li})$ for the vectorization of the channel matrix and $\mathbf{n}_{lk} = \mathrm{vec}(\mathbf{N}_{lk})$ for the noise.
The MMSE estimate of $\mathbf{h}_{lk}$ is \cite[Ch. 3]{bjornson2017massive}
\begin{equation} \label{eq:Uplink_ChannelEstimation}
    \hat{\mathbf{h}}_{lk}= \sqrt{q_{k}\textcolor{black}{\tau_{\mathrm{p}}}}\mathbf{R}_{\mathbf{h}_{lk}}\mathbf{\Psi}_{lk}^{-1}\mathbf{y}_{lk} ,
\end{equation}
where $\mathbf{R}_{\mathbf{h}_{lk}}$ is the covariance of vectorized channel from AP $l$ to user $k$ and $\mathbf{\Psi}_{lk}$ is
\begin{equation} \label{eq:Psi_in_uplinkChannelEstimation}
    \mathbf{\Psi}_{lk} = \mathbb{E} \left\{ \mathbf{y}_{lk} \mathbf{y}_{lk}^{\rm{H}} \right\} =  \tau_{\mathrm{p}} \sum_{{i\in \mathcal{P}_{k}}} q_{i}\mathbf{R}_{\mathbf{h}_{li}}  + \mathbf{I}_{NM} .
\end{equation}
The estimate $\hat{\mathbf{h}}_{lk}\sim \mathcal{N}_{\mathbb{C}}(\mathbf{0},\mathbf{R}_{\hat{\mathbf{h}}_{lk}})$  and the estimation error $\Tilde{\mathbf{h}}_{lk}\sim \mathcal{N}_{\mathbb{C}}(\mathbf{0},\mathbf{C}_{\Tilde{\mathbf{h}}_{lk}})$ are independent random vectors. Here, we have $\mathbf{R}_{\hat{\mathbf{h}}_{lk}} = q_{k}\tau_{\mathrm{p}} \mathbf{R}_{\mathbf{h}_{lk}}\mathbf{\Psi}_{lk}^{-1}\mathbf{R}_{\mathbf{h}_{lk}}$ and $\mathbf{C}_{\Tilde{\mathbf{h}}_{lk}} = \mathbf{R}_{\mathbf{h}_{lk}} - q_{k}\tau_{\mathrm{p}} \mathbf{R}_{\mathbf{h}_{lk}}\mathbf{\Psi}_{lk}^{-1}\mathbf{R}_{\mathbf{h}_{lk}}$ as the covariance matrices of the estimate and the estimation error vectors. \textcolor{black}{The matrix $\Psi_{lk}$ can be obtained based on the long-term statistical properties of the channels, which are typically assumed to be known or estimated in practical systems. In practice, these statistics can be estimated by averaging over many coherence blocks, leveraging the stationary nature of the wireless channel over sufficiently long time intervals. Since these statistics change slowly compared to instantaneous channel realizations, their estimation and sharing among APs is a feasible assumption for cell-free massive MIMO systems.}
\vspace{-3mm}

\subsection{Downlink Data Transmission} \label{sec: Downlink Data Transmission}
\begin{figure}[t!]
\centering
  \includegraphics[width=\linewidth]{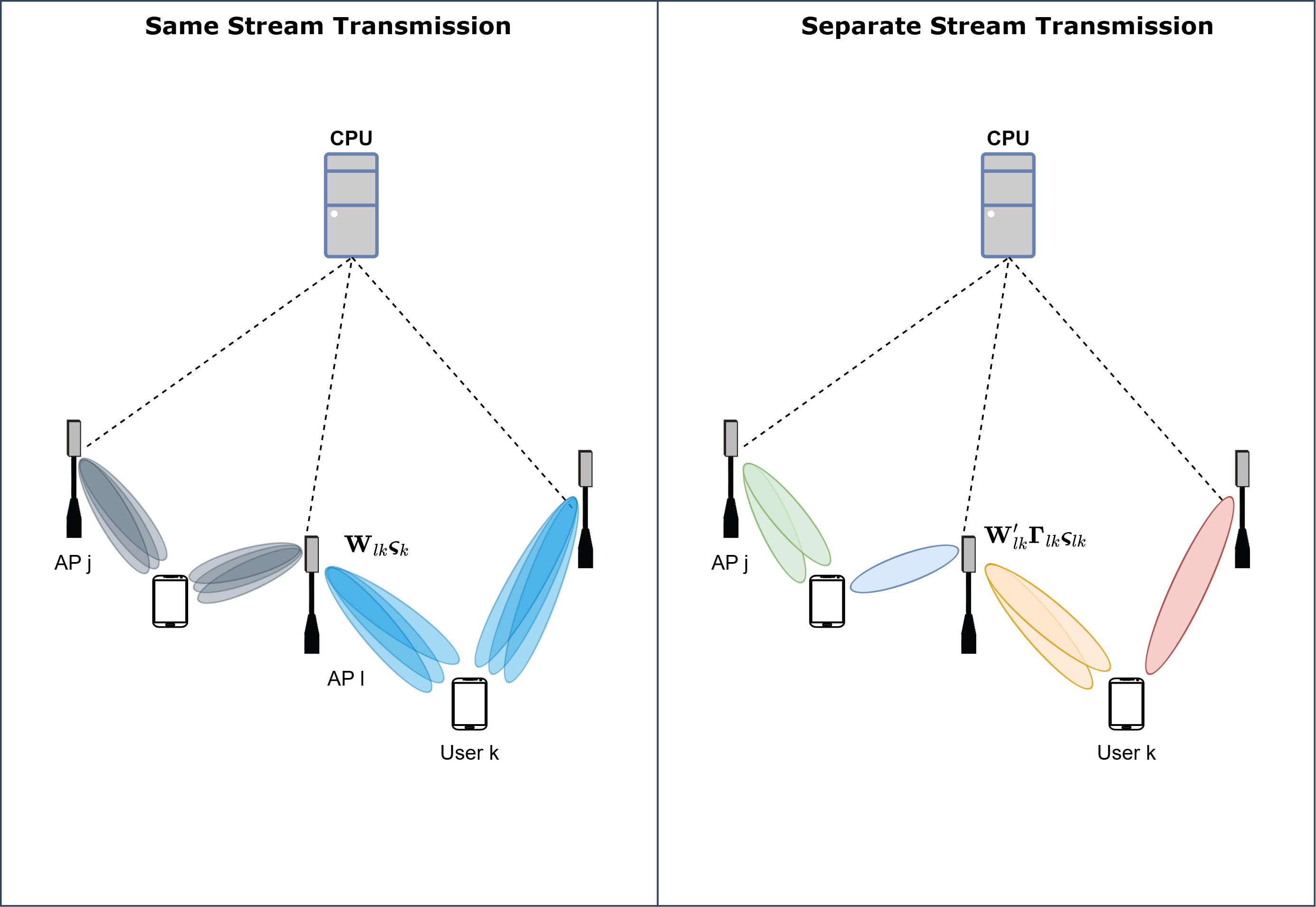} \vspace{-6mm}
  \caption{Illustration of the same stream and the separate stream transmission. Different colors illustrate different data streams. The figure illustrates three streams being received by users as an example, though the number of streams could vary.} \vspace{-3mm}
  \label{fig:the_same_stream_and_separate_stream}
\end{figure}

In this section, we provide system models for downlink data transmission. We will describe two different ways to transmit signals in the downlink. The essence of cell-free massive MIMO with single-antenna users is that the cooperating APs send the same data stream using coherent joint beamforming. When the users have multiple antennas, multiple spatially multiplexed data streams can be sent to each user. Joint transmission then implies that a user is served by multiple APs, but each individual stream can either be transmitted by one or multiple APs. To study these different scenarios, we considered two categories: \textit{same stream transmission} and \textit{separate stream transmission}. The respective system models and notations are detailed in the following subsections. \textcolor{black}{The concepts of same stream transmission and separate stream transmission are illustrated with an example coloring in Fig. \ref{fig:the_same_stream_and_separate_stream}. Transmitted streams from APs are assigned distinct color codes, enabling a visual differentiation between the same stream and separate transmissions.}

\subsubsection{Downlink with Same Stream Transmission}

The most advanced downlink operation involves all APs in the transmission of all data streams, using the centralized operation \cite{demir2021foundations}.
Hence, the APs cooperate in the transmission, share CSI obtained from the uplink channel estimation, and precode their transmitted downlink signals based on it. 

To obtain a convenient notation, we stack all the AP channels to user $k$ in the $LN \times M$ matrix $\mathbf{H}_{k}= [ \mathbf{H}_{1k}^\intercal , \ldots  , \mathbf{H}_{Lk}^\intercal]^\intercal
$. Furthermore, we denote the $LN \times M$ precoder matrix of user $k$ as $\mathbf{W}_{k}= [ \mathbf{W}_{1k}^{\intercal} , \ldots  , \mathbf{W}_{Lk}^{\intercal}]^{\intercal}
 \in \mathbb{C}^{LN \times M} $, where $\mathbf{W}_{lk} \in \mathbb{C}^{N \times M}$ is the part used by AP $l$. Hence, the transmitted data signal $\mathbf{x}_{lk} \in \mathbb{C}^{N}$ from AP $l$ meant for user $k$ is
$\mathbf{x}_{lk}=  \mathbf{W}_{lk} \boldsymbol{\varsigma}_{k}$,
where $\boldsymbol{\varsigma}_{k} \sim \mathcal{N}_{\mathbb{C}}(\mathbf{0},\mathbf{I}_{M})$ is the $M\times 1$ dimensional data signal.
The precoding matrices can be chosen arbitrarily so that 
$\mathbb{E} \left\{ \mathrm{tr}(\mathbf{x}_{l}^{\rm{H}} \mathbf{x}_{l} )\right\} = \sum_{i=1}^{K} \mathbb{E} \left\{ \Vert\mathbf{W}_{li}\Vert^{2}_{F} \right\} \leq \rho_{d}$,
where $\rho_{d}$ is the maximum transmit power and $\mathbf{x}_{l} = \sum_{i=1}^{K} \mathbf{x}_{li} $ is the sum of all transmitted signals from AP $l$. 
We will describe different precoding methods later in Section~\ref{sec: Channel and Precoder Arguments for Option 2  Option 3}.

The received signal at user $k$ is 
\begin{align} \label{downlink_transmitted_signal}
     \nonumber \mathbf{y}_{k} &=  \sum_{j=1}^{L}  \mathbf{H}_{jk}^{\rm{H}} \mathbf{W}_{jk} \boldsymbol{\varsigma}_{k} + \sum_{i=1, i \neq k}^{K} \sum_{j=1}^{L}  \mathbf{H}_{jk}^{\rm{H}} \mathbf{W}_{ji} \boldsymbol{\varsigma}_{i} +\mathbf{n}_{k} \\
     &=   \mathbf{H}_{k}^{\rm{H}} \mathbf{W}_{k} \boldsymbol{\varsigma}_{k} + \sum_{i=1, i \neq k}^{K}  \mathbf{H}_{k}^{\rm{H}} \mathbf{W}_{i} \boldsymbol{\varsigma}_{i} +\mathbf{n}_{k},
\end{align}
where $\mathbf{n}_{k} \sim \mathcal{N}_{\mathbb{C}}(\mathbf{0},\mathbf{I}_{M})$ is the normalized noise vector. The channel and precoding matrices appear in the form $\mathbf{B}_{ki}=\mathbf{H}_{k}^{\rm{H}} \mathbf{W}_{i} \in \mathbb{C}^{M \times M}$ in \eqref{downlink_transmitted_signal}. We will use this notation for the effective channel after precoding in the remainder of this paper when considering the same stream transmission.

For a given realization of the effective channels, created by the channel realizations and precoding scheme, the achievable SE varies depending on the available CSI at the receiver. We will provide expressions for three different cases in Section~\ref{sec: SE Results for Option 1} and compare them numerically in Section~\ref{Numerical_Results}.

\subsubsection{Downlink with Separate Stream Transmission}

In practice, it might be desirable for each data stream to only be transmitted from one AP because then \textcolor{black}{the} fronthaul signaling related to data will not grow with the number of APs. In this case, we also alleviate the need for phase-synchronization between the APs since there is no beamforming of the same signals that must superimpose coherently over the air.

In this subsection, we provide the essential notation for this decentralized method, where each AP sends a different independent stream. As the maximum achievable rank of the channel $\mathbf{H}_k$ is $M$, no more than $M$ streams should be sent from the different APs.
Moreover, at most $M$ of the $L$ APs are transmitting anything to a specific user.


All the serving APs use the uplink channel estimates and form the transmitted signal with a precoder matrix $\mathbf{W}_{jk} = \mathbf{W}'_{jk}\boldsymbol{\Gamma}_{jk}$ between AP $j$ and user $k$, where $\mathbf{W}'_{jk } \in \mathbb{C}^{N \times M}$. To decide which data streams are sent from an AP, we introduce the selection matrix $\boldsymbol{\Gamma}_{jk}\in \mathbb{C}^{M \times M}$.
This is a binary diagonal matrix with the number of $1$s equalling the number of streams that AP $j$ sends to user $k$.
All matrix entries are zero if AP $j$ doesn't transmit any streams to user $k$. The reason for this choice is to keep the dimensions of precoders and streams the same as in the previous section. 
The positions of the $1$s decide which of the (up to) $M$ streams are sent from AP $j$, since the selection matrix $\boldsymbol{\Gamma}_{jk}$ selects which column vectors of the precoder matrix $\mathbf{W}'_{jk}$ are used. 

Note that self-multiplication of a selection matrix gives $\boldsymbol{\Gamma}_{jk}\boldsymbol{\Gamma}_{jk}=\boldsymbol{\Gamma}_{jk}$, which allows us to group them or separate them when necessary. Furthermore, these matrices are chosen so that $\boldsymbol{\Gamma}_{jk}\boldsymbol{\Gamma}_{j'k}=\mathbf{0}_{M}$ for $j\neq j'$ and fixed $k$ to avoid sending the same stream from multiple APs. 
As only selected columns are non-zero in a precoder matrix $\mathbf{W}_{jk}$ computed in this way, we need specific methods for selecting $\mathbf{W}'_{jk}$. These will be provided in Section \ref{sec: Channel and Precoder Arguments for Option 2  Option 3}. 
By stacking the $\boldsymbol{\Gamma}_{jk}$ matrices, we obtain  $\boldsymbol{\Gamma}_{k}= [ \boldsymbol{\Gamma}_{1k}^\intercal , \ldots  , \boldsymbol{\Gamma}_{Lk}^\intercal]^\intercal
\in \mathbb{C}^{LM \times M}$. Moreover, we stack all the $\mathbf{W}'_{lk}$ matrices block diagonally to obtain $\mathbf{W}'_{k}=\mathrm{diag}(\mathbf{W}'_{1k},\ldots,\mathbf{W}'_{Lk}) \in \mathbb{C}^{LN \times LM}$. Therefore, the total precoder matrix has the form \textcolor{black}{$\mathbf{W}_{k}= \mathbf{W}'_{k}\boldsymbol{\Gamma}_{k}=[ (\mathbf{W}_{1k}\boldsymbol{\Gamma}_{1k})^\intercal , \ldots  , (\mathbf{W}_{Lk}\boldsymbol{\Gamma}_{Lk})^\intercal]^{\intercal}  \in \mathbb{C}^{LN \times M}
$.}

The $N \times 1$ transmitted signal from AP $l$ to user $k$ is $\mathbf{x}_{lk}= \mathbf{W}_{lk} \boldsymbol{\varsigma}_{k} = \mathbf{W}'_{lk}\boldsymbol{\Gamma}_{lk} \boldsymbol{\varsigma}_{k}$, where $\boldsymbol{\varsigma}_{k} = [\varsigma_{1k},\hdots,\varsigma_{Mk}]^\intercal$ is the vector of streams. 
The received signal at user $k$ is 
\begin{align} \label{downlink_transmitted_signal_2}
     \nonumber \mathbf{y}_{k} &=  \sum_{j = 1}^{L}  \mathbf{H}_{jk}^{H} \mathbf{W}'_{jk} \boldsymbol{\Gamma}_{jk} \boldsymbol{\varsigma}_{k} + \sum_{i=1, i \neq k}^{K} \sum_{j = 1}^{L}  \mathbf{H}_{jk}^{H} \mathbf{W}'_{ji} \boldsymbol{\Gamma}_{ji} \boldsymbol{\varsigma}_{i}+\mathbf{n}_{k} \\ 
     &=   \mathbf{B}'_{kk} \boldsymbol{\Gamma}_{k}\boldsymbol{\varsigma}_{k} +  \sum_{i=1, i \neq k}^{K}  \mathbf{B}'_{ki} \boldsymbol{\Gamma}_{i}\boldsymbol{\varsigma}_{i} +\mathbf{n}_{k} ,
\end{align}
where $\mathbf{n}_{k} \sim \mathcal{N}_{\mathbb{C}}(\mathbf{0},\mathbf{I}_{M})$ is the normalized noise vector.
Here, $\mathbf{B}'_{ki}=\mathbf{H}_{k}^{H} \mathbf{W}'_{i}$ denotes the  $M \times LM$ dimensional matrices containing $\mathbf{H}_{jk}^{H} \mathbf{W}'_{ji}$ terms stacked as a row matrix. With the selection matrices, the effective channels are $\mathbf{B}'_{ki}\boldsymbol{\Gamma}_{i}\in \mathbb{C}^{M \times M}$. We will use this notation to highlight how the selection matrices affect the SE expressions derived in Section~\ref{sec: SE Results for Option 2}.

\section{Downlink Achievable SE Results} \label{sec: SE Results for Option 1}
The described uplink channel estimation procedure provides the APs with CSI and determines which precoding schemes they can use. However, it does not provide the users with any channel knowledge. The available CSI at the receiver determines what receiver processing can be used and eventually how high the SE becomes. Given this relationship, in this section, we will investigate the achievable SE with three kinds of CSI availability at the receivers.

\subsection{Hardening Bound without CSI at the Receiver} \label{Hardening Bound Without CSI at the Receiver}

When the receiver lacks CSI regarding the instantaneous channel realizations, the SE can be lower bounded using the hardening bound \cite{demir2021foundations}. This method is conventionally used for single-antenna users, but the multi-antenna extension was considered in \cite{buzzi2019user,zhou2021sum,li2016massive, mai2020downlink, wang2022uplink,sutton2021hardening}. 
The main idea is that user $k$ knows the statistics of its effective channel $\mathbf{B}_{kk} = \mathbf{H}_{k}^{\rm{H}} \mathbf{W}_{k}$, such as its mean $\bar{\mathbf{B}}_{kk} = \mathbb{E}\left\{   \mathbf{B}_{kk} \right\}$, and the precoder matrix has been selected based on the CSI to make the variations around the mean small. We rewrite the received signal in \eqref{downlink_transmitted_signal} as
\begin{equation} \label{eq:rewritten-hardening-bound}
     \mathbf{y}_{k} =  \bar{\mathbf{B}}_{kk} \boldsymbol{\varsigma}_{k} + \underbrace{(\mathbf{B}_{kk}-\bar{\mathbf{B}}_{kk}) \boldsymbol{\varsigma}_{k}+ \sum_{i=1, i \neq k}^{K} \mathbf{B}_{ki} \boldsymbol{\varsigma}_{i} +\mathbf{n}_{k}}_{\triangleq \mathbf{n}'_{k}},
\end{equation}
where $\mathbf{n}'_{k} \in \mathbb{C}^{M \times 1}$ denotes the sum of noise, the desired signal received over the unknown channel component (i.e., the deviation from the mean), and inter-user interference. 
This term is uncorrelated with the first term $\bar{\mathbf{B}}_{kk} \boldsymbol{\varsigma}_{k}$ in \eqref{eq:rewritten-hardening-bound}, but it is spatially colored with the covariance matrix 
\begin{equation}
     \mathbf{\Xi}_{k} =  \mathbb{E}\left\{   \mathbf{n}'_{k} \mathbf{n}'^{\rm{H}}_{k} \right\}. 
\end{equation}

\begin{lemma} \label{first_lemma}
In the absence of instantaneous CSI at the user side, an achievable SE at user $k$ is
\begin{equation} \label{eq: first_lemma}
    \mathrm{SE}_{k}^\mathrm{noCSI}=\left(1 - \frac{\tau_{\mathrm{p}}}{\tau_{\mathrm{c}}}\right) \log_{2} \left|\mathbf{I}_{M} + \bar{\mathbf{B}}_{kk}^{\rm{H}} \mathbf{\Xi}_{k}^{-1} \bar{\mathbf{B}}_{kk}\right|,
\end{equation}
where $| \cdot |$ denotes the determinant.
\end{lemma}


The proof of this lemma builds on interpreting \eqref{eq:rewritten-hardening-bound} as a point-to-point MIMO system with the deterministic channel matrix $\bar{\mathbf{B}}_{kk}$ and uncorrelated additive noise with the covariance matrix $\mathbf{\Xi}_{k} \in \mathbb{C}^{M \times M}$, \textcolor{black}{which can be obtained by using sample averages}. The SE expression is then obtained as a lower bound on the channel capacity by utilizing the worst-case uncorrelated additive noise term \cite{Hassibi2003a}.
The single-antenna version of Lemma~\ref{first_lemma} is called the hardening bound \cite{demir2021foundations} since the effective channel realization $\mathbf{B}_{kk}$ is close to its mean  $\bar{\mathbf{B}}_{kk}$ (in relative terms) when $M=1$, $N$ is large, and the precoder is selected as MR, ZF, or a filter of similar structure. This phenomenon is called channel hardening.
However, as the effective channel is a matrix instead of a scalar when $M>1$, the same kind of channel hardening is not achieved herein since each column of the precoder matrix can only achieve hardening with respect to one row in the channel matrix. The resulting performance loss is demonstrated later in this paper.

To explain the insufficient channel hardening analytically, we will provide a simple example with MR precoding based on perfect CSI where the precoder matrix is selected as $\mathbf{W}_{lk}=\mathbf{H}_{lk}$.
We consider i.i.d. Rayleigh fading with $\mathbf{R}_{lk} = \mathbf{I}_{NM}$ \textcolor{black}{only for this example case}.
We further assume there is only one AP and one user, so the related indices can be dropped.
The effective channel $\mathbf{B}=\mathbf{H}^{\rm{H}}\mathbf{W}$ has the mean $\bar{\mathbf{B}} = \mathbb{E}\{ \mathbf{B} \} = N \mathbf{I}_M$. 
The matrix $\mathbf{B}-\bar{\mathbf{B}}$ represents the true effective channel's deviation from its mean.
It is straightforward to show that all its entries are independent and have variance $N$; thus, the standard deviation is $\sqrt{N}$.

When $N$ is large, the diagonal entries of $\bar{\mathbf{B}}$ are much larger than the diagonal entries of $\mathbf{B}-\bar{\mathbf{B}}$. This is the essence of channel hardening: we can neglect the diagonal of $\mathbf{B}-\bar{\mathbf{B}}$ since the corresponding entries of $\bar{\mathbf{B}}$ are roughly $N/\sqrt{N} = \sqrt{N}$ times larger.
By contrast, the off-diagonal entries of $\bar{\mathbf{B}}$ are zero, while the corresponding entries of $\mathbf{B}-\bar{\mathbf{B}}$ have a standard deviation of $\sqrt{N}$ and cannot be neglected when $N$ is large.

We conclude that only the diagonal entries of $\mathbf{B}_{kk}$ harden, while the off-diagonal entries that represent the inter-stream interference do not.
There are no off-diagonal entries when $M=1$, so channel hardening holds in that special case (as shown in \cite{ngo2017no,demir2021foundations}), while most entries do not harden when $M>1$. Therefore, the hardening bound is loose when $M>1$, which calls for developing a methodology for acquiring CSI at the receiving user using downlink pilots.\footnote{The pilot-free downlink CSI acquisition method proposed in \cite{ngo2017no} cannot be extended to the multi-antenna case because it relies on the proposal that the precoder matrix is selected to make the entries of $\mathbf{B}$ (approximately) real-valued and positive. This can only be guaranteed for $M$ entries of $\mathbf{B}$, while the remaining $M^2-M$ (typically off-diagonal entries) cannot be estimated in this way.}

\textcolor{black}{The absence of channel hardening in i.i.d. Rayleigh fading for $ M > 1 $ inherently implies that correlated Rayleigh fading will also fail to exhibit hardening. When channel correlation is introduced, the covariance structure $\mathbf{R}_{lk}$ modifies the statistical properties of $\mathbf{B}$, but the fundamental issue persists. The off-diagonal terms remain zero-mean with variances dependent on the correlation matrix. Correlation often aggravates the variability of these terms by introducing structured dependencies between channel coefficients, amplifying their deviations relative to the mean. }

\subsection{Capacity Bound with Perfect CSI at the Receiver}

Next, we consider the ideal case with perfect knowledge of the effective channels at the receiver. The CSI is obtained in a genie-aided way, which makes this case an upper bound that we can use as a benchmark for other practical methods. We write the downlink received signal in \eqref{downlink_transmitted_signal} as
\begin{equation} 
\begin{split}
     \mathbf{y}_{k} =   \mathbf{B}_{kk} \boldsymbol{\varsigma}_{k} + \mathbf{n}''_{k},
\end{split}
\end{equation}
where $\mathbf{n}''_{k} = \sum_{i=1, i \neq k}^{K}  \mathbf{H}_{k}^{\rm{H}} \mathbf{W}_{i} \boldsymbol{\varsigma}_{i} +\mathbf{n}_{k}$ is composed of the independent noise and interference terms.

\begin{lemma} \label{second_lemma}
With perfect CSI, an achievable SE at user $k$ is 
\begin{equation} \label{Perfect_CSI_SE}
    \mathrm{SE}_{k}^\mathrm{fullCSI} = \left(1 - \frac{\tau_{\mathrm{p}}}{\tau_{\mathrm{c}}}\right) \mathbb{E}\left\{  \log_{2} \left|  \mathbf{I}_{M} +  \mathbf{B}_{kk}^{\rm{H}}\Tilde{\mathbf{\Xi}}_{k}^{-1} \mathbf{B}_{kk}  \right| \right\},
\end{equation}
where the covariance matrix $\Tilde{\mathbf{\Xi}}_{k} \in \mathbb{C}^{M \times M}$ of $\mathbf{n}''_{k} \in \mathbb{C}^{M \times 1}$ is
\begin{align}
     \Tilde{\mathbf{\Xi}}_{k} =        \mathbb{E}\left\{   \mathbf{n}''_{k} \mathbf{n}''^{\rm{H}}_{k} \right\}
     \nonumber = \sum_{i=1, i \neq k}^{K}  \mathbf{B}_{ki} \mathbf{B}_{ki}^{\rm{H}} +\mathbf{I}_{M}.
\end{align}
\end{lemma}

This result is proved by treating the term $\mathbf{n}''_{k}$ as worst-case Gaussian noise, whitening the noise, and then stating the ergodic SE with perfect CSI at the receiver.
We will later show by simulations that there is a large gap between the SE with perfect CSI in Lemma~\ref{second_lemma} and without CSI at the receiver in Lemma~\ref{first_lemma}. Therefore, we will analyze downlink channel estimation as a means to improve the SE.


\subsection{Estimation of the Effective Downlink Channel}

A viable way to provide the receiver with CSI is to send downlink pilots that are precoded just as the data, so that user~$k$ can estimate the effective channel $\mathbf{B}_{kk}$. Since the effective channel takes random non-Gaussian realizations from a stationary distribution, a suitable estimation method is LMMSE estimation \cite[Ch. 12]{kay1993fundamentals}. To enable the estimation, the APs jointly send orthonormal pilot sequences using the selected precoding. The pilot matrix assigned to user $k$ is $\boldsymbol{\Phi}_{k} \in \mathbb{C}^{\tau_{\mathrm{p}} \times M}$. Again, $\mathcal{P}_{k}$ is the set of users that use the same pilot matrix as user $k$. The pilot matrices are orthonormal, that is, $\boldsymbol{\Phi}_{i}^{\rm{H}} \boldsymbol{\Phi}_{i'} = \mathbf{I}_{M}$ if $i , i'\in \mathcal{P}_{k}$ and $\boldsymbol{\Phi}_{i}^{\rm{H}} \boldsymbol{\Phi}_{i'} = \mathbf{0}$ otherwise.\footnote{Note that we use the same pilot length $\tau_{\mathrm{p}}$ as in the uplink, but in principle, one could use a different length.} By sending these pilots over the channel in \eqref{downlink_transmitted_signal}, the received signal at user $k$ becomes

\begin{equation}
    \tilde{\mathbf{Y}}_{k}^{\text{Pilot}}=\sum_{i=1}^{K} \sqrt{q_i \tau_{\mathrm{p}}} \mathbf{B}_{ki} \boldsymbol{\Phi}_{i}^{\rm{H}} + \mathbf{N}^{\text{Pilot}}_{k},
\end{equation}
\textcolor{black}{where $\mathbf{N}^{\text{Pilot}}_{k}$ is the noise matrix at user $k$ for pilot transmission with i.i.d. $\mathcal{N}_{\mathbb{C}}(0,1)$-entries.} By correlating with the dedicated pilot for user $k$, we obtain
\begin{align}   
    \tilde{\mathbf{Y}}_{k} &=\tilde{\mathbf{Y}}_{k}^{\text{Pilot}} \boldsymbol{\Phi}_{k}=\sum_{i=1}^{K} \sqrt{q_i \tau_{\mathrm{p}}} \mathbf{B}_{ki} \boldsymbol{\Phi}_{i}^{\rm{H}}  \boldsymbol{\Phi}_{k} + \mathbf{N}_{k} \nonumber \\ \label{downlink_channel_estimate0}
     &= \sqrt{q_k \tau_{\mathrm{p}}} \mathbf{B}_{kk} + \sum_{{i \in \mathcal{P}_k \setminus \{ k\}}} \sqrt{q_i \tau_{\mathrm{p}}} \mathbf{B}_{ki} + \mathbf{N}_{k},
\end{align}
where $\mathbf{N}_{k}=\mathbf{N}^{\text{Pilot}}_{k}\boldsymbol{\Phi}_{k}$ is the noise after correlation with the pilot matrix and only interference from pilot-sharing users remain.
The entries of $\mathbf{B}_{ki}$ matrices are statistically correlated. To describe the correlation using covariance matrices, we first need to 
vectorize the observation equation in \eqref{downlink_channel_estimate0}:
\begin{equation} \label{downlink_channel_estimate}
\begin{split}
    \mathrm{vec}(\tilde{\mathbf{Y}}_{k}) &= \sqrt{q_k \tau_{\mathrm{p}}} \mathrm{vec}(\mathbf{B}_{kk}) + \!\!\!\! \sum_{{i \in \mathcal{P}_k \setminus \{ k\}}} \!\!\!\sqrt{q_i \tau_{\mathrm{p}}} \mathrm{vec}(\mathbf{B}_{ki}) + \mathbf{n}_{k} ,
\end{split}
\end{equation}
where $\mathbf{n}_{k} = \mathrm{vec}(\mathbf{N}_{k})$.
We will let $\mathbf{b}_{kk} = \mathrm{vec}(\mathbf{B}_{kk})$ and $\mathbf{\tilde{y}}_{k} = \mathrm{vec}(\tilde{\mathbf{Y}}_{k})$ denote the vectorized effective channel and received signal, respectively. The LMMSE estimate of $\mathbf{b}_{kk}$ is \cite[Ch. 12]{kay1993fundamentals}
\begin{equation} \label{eq:LMMSE_estimator}
    \hat{\mathbf{b}}_{kk} =   \mathbb{E}\left\{ \mathbf{b}_{kk} \right\} +  \mathbf{C}_{\mathbf{b}_{kk} \mathbf{\tilde{y}}_{k}}\mathbf{C}_{\tilde{\mathbf{y}}_{k}}^{-1}    (\mathbf{\tilde{y}}_{k} - \mathbb{E}\left\{\mathbf{\tilde{y}}_{k} \right\} ),
\end{equation}
where the covariance matrices $\mathbf{C}_{\mathbf{b}_{kk} \mathbf{\tilde{y}}_{k}},\mathbf{C}_{\mathbf{b}_{kk}}, \mathbf{C}_{\tilde{\mathbf{y}}_{k}}$ are cross-covariance of effective channel and observation, the covariance of effective channel and the covariance of received signal respectively. \textcolor{black}{To compute the LMMSE estimate of $\mathbf{b}_{kk}$, the required expectations, such as $\mathbb{E}\left\{ \mathbf{b}_{kk} \right\}$, $\mathbb{E}\left\{\mathbf{\tilde{y}}_{k} \right\}$, and the covariance matrices are assumed to be known. In practice, they could be estimated by averaging over multiple channel realizations or observations, assuming the underlying statistics remain consistent over time.} The estimation error covariance of this LMMSE estimator is 
\begin{equation}
\mathbf{C}_{\tilde{\mathbf{b}}_{kk}} = \mathbf{C}_{\mathbf{b}_{kk}} - \mathbf{C}_{\mathbf{b}_{kk} \mathbf{\tilde{y}}_{k}} \mathbf{C}_{\mathbf{\tilde{y}}_{k}}^{-1}
\mathbf{C}_{\mathbf{b}_{kk} \mathbf{\tilde{y}}_{k}}
\end{equation}
and the total MSE is 
$\text{MSE} = \text{tr}(\mathbf{C}_{\tilde{\mathbf{b}}_{kk}})$.
\vspace{-3mm}
\subsection{Capacity Bound with Estimated Channel at the Receiver} \label{Capacity Bound with Estimate Channel at the Receiver}

We will now derive an achievable SE expression for the case when the receiver uses the proposed LMMSE estimator to acquire the effective channel matrix. Since the estimate $\hat{\mathbf{b}}_{kk}$ and estimation error $\mathbf{b}_{kk}-\hat{\mathbf{b}}_{kk}$ are non-Gaussian, they are uncorrelated but statistically dependent which rules out the use of classical capacity bounds. We will derive a novel capacity lower bound for this scenario by following the use-and-then-forget (UatF) approach, where the receiver uses its CSI for receive combining, and then we write the effective channel as a deterministic quantity. Applying a general receive combining matrix $\mathbf{U}_{k} \in \mathbb{C}^{M \times M}$ to \eqref{downlink_transmitted_signal}, we obtain

\begin{align} 
      &\tilde{\mathbf{y}}_{k} =  \mathbf{U}_{k}^{\rm{H}} \mathbf{y}_{k} =  \mathbf{U}_{k}^{\rm{H}}  \mathbf{B}_{kk} \boldsymbol{\varsigma}_{k} + \mathbf{U}_{k}^{\rm{H}} \sum_{i=1, i \neq k}^{K} \mathbf{B}_{ki} \boldsymbol{\varsigma}_{i} + \mathbf{U}_{k}^{\rm{H}} \mathbf{n}_{k} \nonumber \\
      &=  \underbrace{\mathbf{U}_{k}^{\rm{H}}\hat{\mathbf{B}}_{kk}}_{\triangleq \mathbf{E}_{kk}} \boldsymbol{\varsigma}_{k} + \mathbf{U}_{k}^{\rm{H}}  \Big(  \tilde{\mathbf{B}}_{kk} \boldsymbol{\varsigma}_{k} + \sum_{i=1, i \neq k}^{K} \mathbf{B}_{ki} \boldsymbol{\varsigma}_{i} +  \mathbf{n}_{k} \Big)  \label{eq:observation_equation_after_general_combiner}\\
      &=   \bar{\mathbf{E}}_{kk} \boldsymbol{\varsigma}_{k} + (\mathbf{E}_{kk} - \bar{\mathbf{E}}_{kk})\boldsymbol{\varsigma}_{k}  \nonumber\\& \hspace{4em}+ \mathbf{U}_{k}^{\rm{H}}  \Big(  \tilde{\mathbf{B}}_{kk} \boldsymbol{\varsigma}_{k}+ \sum_{i=1, i \neq k}^{K} \mathbf{B}_{ki} \boldsymbol{\varsigma}_{i} +  \mathbf{n}_{k} \Big)\nonumber\\
      &=  \bar{\mathbf{E}}_{kk} \boldsymbol{\varsigma}_{k} + \grave{\mathbf{n}}_{k}, \label{eq:observation_equation_after_general_combiner_last}
\end{align}
where \textcolor{black}{$\hat{\mathbf{B}}_{ki}$ denotes the LMMSE estimate of $\mathbf{B}_{ki}$ in matrix form and the effective channel estimation error is $\tilde{\mathbf{B}}_{ki}=\mathbf{B}_{ki}-\hat{\mathbf{B}}_{ki}$.} $\mathbf{E}_{kk} \in \mathbb{C}^{M \times M}$ denotes the effective channel after combining and $\bar{\mathbf{E}}_{kk}$  the mean of it. The term $\grave{\mathbf{n}}_{k} \in \mathbb{C}^{M \times 1}$ contains the unknown part of the effective channel, interference, mismatch between the effective channel and its mean, and noise after combining. By realizing that the effective channel $\bar{\mathbf{E}}_{kk}$ in front of the streams in \eqref{eq:observation_equation_after_general_combiner_last} is now deterministic, we can write the following SE result.

\begin{theorem}\label{theorem:mainresult_general_combiner}
When downlink pilots, the LMMSE estimator, and a general combiner are used, an achievable SE at user $k$ is
\begin{equation} \label{eq:lemma_main_result_general_combiner}
     \mathrm{SE}_{k}^\mathrm{pilots} = \left(1 - \frac{2\tau_{\mathrm{p}}}{\tau_{\mathrm{c}}}\right) \log_{2} \left| \mathbf{I}_{M} + \bar{\mathbf{E}}_{kk}^{\rm{H}}\mathbf{C}_{\grave{\mathbf{n}}_{k}}^{-1}\bar{\mathbf{E}}_{kk}  \right| , 
\end{equation}
where $\mathbf{C}_{\grave{\mathbf{n}}_{k}} \in \mathbb{C}^{M \times M}$ is the covariance of the term defined in \eqref{eq:observation_equation_after_general_combiner_last}.
\end{theorem}
\begin{IEEEproof}
    The proof is given in Appendix \ref{Appendix: Proof of main result general combiner}.
\end{IEEEproof}
\textcolor{black}{We note that the SE relation captures both the data rate during data transmission and the loss from the pilot overhead. Resulting effects of this and the improvement in SE will be shown in Section \ref{Numerical_Results}.}
There are multiple ways that the user can select the receive combining matrix.
One option is the MMSE combiner, which can be obtained by writing the MSE relation $\mathbb{E}\left\{ |\mathbf{U}_{k}^{\rm{H}} \mathbf{y}_{k}-\boldsymbol{\varsigma}_{k}|^{2} \big| \hat{\mathbf{B}}_{ki} \; \mathrm{for} \; i=1,\hdots, K \right\}$ and finding the filter that minimizes it. The MMSE combiner for user $k$ can be written as
\begin{equation} \label{eq:combiner for the effective channel}
\mathbf{U}_{k} = \left(\sum_{i=1}^{K}(\hat{\mathbf{B}}_{ki} \hat{\mathbf{B}}_{ki}^{\rm{H}} + \mathbf{C}_{\tilde{\mathbf{B}}_{ki}}) + \mathbf{I}_{M} \right)^{-1} \hat{\mathbf{B}}_{kk}. 
\end{equation}
Here, $\mathbf{C}_{\tilde{\mathbf{B}}_{ki}} = \mathbb{E}\left\{\tilde{\mathbf{B}}_{ki} \tilde{\mathbf{B}}_{ki}^{\rm{H}} \right\}$ denotes the covariance of the downlink channel estimation error. 

We can obtain a simpler SE expression by selecting a receive combining matrix that makes the channel $\mathbf{E}_{kk}$ deterministic, so the deviation term $\mathbf{E}_{kk} - \bar{\mathbf{E}}_{kk}$ in \eqref{eq:observation_equation_after_general_combiner} becomes zero.
This is achieved by the ZF combining matrix  $\mathbf{U}_{k}^{\rm{H}} = (\hat{\mathbf{B}}_{kk}^{\rm{H}} \hat{\mathbf{B}}_{kk})^{-1} \hat{\mathbf{B}}_{kk}^{\rm{H}}$.\footnote{Since $\hat{\mathbf{B}}_{kk}$ is a square matrix, the ZF combiner is equivalent to $\hat{\mathbf{B}}_{kk}^{-1}$. However, we keep the definition as here for the cases where fewer streams are sent, leading to a decrease in the rank.} By expressing the unknown part of the channel as $\tilde{\mathbf{B}}_{kk} = \mathbf{B}_{kk} - \hat{\mathbf{B}}_{kk}$, \eqref{eq:observation_equation_after_general_combiner} can be expressed as
\begin{equation} \label{eq:observation_equation_afterZFcombiner}
\begin{split}
      \tilde{\mathbf{y}}_{k} &=  (\hat{\mathbf{B}}_{kk}^{\rm{H}} \hat{\mathbf{B}}_{kk})^{-1} \hat{\mathbf{B}}_{kk}^{\rm{H}}\hat{\mathbf{B}}_{kk} \boldsymbol{\varsigma}_{k} + \\ 
      &  \quad (\hat{\mathbf{B}}_{kk}^{\rm{H}} \hat{\mathbf{B}}_{kk})^{-1} \hat{\mathbf{B}}_{kk}^{\rm{H}} \Big(  \tilde{\mathbf{B}}_{kk} \boldsymbol{\varsigma}_{k} + \sum_{i=1, i \neq k}^{K} \mathbf{B}_{ki} \boldsymbol{\varsigma}_{i} +  \mathbf{n}_{k} \Big) \\
      &=   \boldsymbol{\varsigma}_{k} + \underbrace{(\hat{\mathbf{B}}_{kk}^{\rm{H}} \hat{\mathbf{B}}_{kk})^{-1} \hat{\mathbf{B}}_{kk}^{\rm{H}} \Big(  \tilde{\mathbf{B}}_{kk} \boldsymbol{\varsigma}_{k} + \sum_{i=1, i \neq k}^{K} \mathbf{B}_{ki} \boldsymbol{\varsigma}_{i} +  \mathbf{n}_{k} \Big)}_{\triangleq \grave{\mathbf{n}}_{k}}.
\end{split}
\end{equation}
We notice that this is the transmitted signal $\boldsymbol{\varsigma}_{k}$ plus the uncorrelated interference and noise term $\grave{\mathbf{n}}_{k} \in \mathbb{C}^{M \times 1}$. Hence, the effective channel has been equalized to an identity matrix and we can simplify the result in Theorem~\ref{theorem:mainresult_general_combiner} as follows.

\begin{corollary} \label{th:mainresult}
When downlink pilots, the LMMSE estimator, and ZF combining are utilized, an achievable SE at user $k$ is
\begin{equation} \label{theorem_1_main_result}
     \mathrm{SE}_{k}^\mathrm{pilotsZF} = \left(1 - \frac{2\tau_{\mathrm{p}}}{\tau_{\mathrm{c}}}\right) \log_{2} \left| \mathbf{I}_{M} + \mathbf{C}_{\grave{\mathbf{n}}_{k}}^{-1}  \right|, 
\end{equation}
where $\mathbf{C}_{\grave{\mathbf{n}}_{k}}  \in \mathbb{C}^{M \times M}$ is the covariance of the term defined in \eqref{eq:observation_equation_afterZFcombiner}.
\end{corollary}
\begin{IEEEproof}
    The proof is similar to the one in Appendix \ref{Appendix: Proof of main result general combiner} and is omitted for brevity.
\end{IEEEproof}

When comparing the new SE expressions in Theorem~\ref{theorem:mainresult_general_combiner} and Corollary~\ref{th:mainresult} with Lemma \ref{first_lemma}, we notice that the new pre-log factor is smaller since it compensates for the fact that pilots are transmitted in both uplink and downlink. Nevertheless, we expect the SE values to be larger and the benefit comes from the matrix expression inside the determinant, where Theorem~\ref{theorem:mainresult_general_combiner} will give a larger SE since we use the downlink channel estimate to equalize the received signal as in \eqref{eq:observation_equation_afterZFcombiner}. 

The ZF combiner yields an easier-to-interpret SE expression than the MMSE combiner, but we anyway expect the latter to give larger SE values as it finds a better balance between signal equalization and interference suppression.
A comparison between these different combining schemes will be given in Section~\ref{Numerical_Results}.


\section{SE Results for Separate Stream Transmission} \label{sec: SE Results for Option 2}
In this section, we will determine the SE relations for separate stream transmission. The relations have the same structure as in the former section, but can be rewritten to showcase the inner structure of the SE expressions.
We note that the precoder structure must also change to reflect this change in SE expression. We will consider that in Section~\ref{sec: Channel and Precoder Arguments for Option 2  Option 3}, while the achievable SEs will be compared numerically in Section \ref{Numerical_Results}. 

\subsection{Downlink Achievable SE without CSI at the Receiver} 

If the receiver lacks CSI, we can obtain an SE with the hardening bound similar to Lemma \ref{first_lemma}. 
To this end, we first write the received signal in \eqref{downlink_transmitted_signal_2} as
\begin{equation} \label{eq:rewritten-hardening-bound_opt2}
     \mathbf{y}_{k} =  \bar{\mathbf{B}}'_{kk} \boldsymbol{\Gamma}_{k}\boldsymbol{\varsigma}_{k} + \underbrace{(\mathbf{B}'_{kk}-\bar{\mathbf{B}}'_{kk}) \boldsymbol{\Gamma}_{k}\boldsymbol{\varsigma}_{k}+ \hspace{-0.5em}\sum_{i=1, i \neq k}^{K} \mathbf{B}'_{ki} \boldsymbol{\Gamma}_{i}\boldsymbol{\varsigma}_{i} +\mathbf{n}_{k}}_{\triangleq \mathbf{n}'_{k}},
\end{equation}
where $\bar{\mathbf{B}}'_{kk} = \mathbb{E}\left\{   \mathbf{B}'_{kk} \right\} $ and $\mathbf{n}'_{k} $ denotes the sum of noise, the desired signal received over the unknown channel component, and inter-user interference. 
This spatially colored noise term is uncorrelated with the first term $\bar{\mathbf{B}}'_{kk} \boldsymbol{\Gamma}_{k} \boldsymbol{\varsigma}_{k}$ in \eqref{eq:rewritten-hardening-bound_opt2}, and it has the covariance matrix 
\begin{equation}
     \mathbf{\Xi}_{k} =  \mathbb{E}\left\{   \mathbf{n}'_{k} \mathbf{n}'^{\rm{H}}_{k} \right\}. 
\end{equation}
We can restate Lemma \ref{first_lemma} for this case using the selection matrices as
\begin{equation} \label{eq: first_lemma_opt2_variation}
    \mathrm{SE}_{k}^\mathrm{noCSI}=\left(1 - \frac{\tau_{\mathrm{p}}}{\tau_{\mathrm{c}}}\right) \log_{2} \left|\mathbf{I}_{M} + \boldsymbol{\Gamma}_{k}^{\rm{H}}\bar{\mathbf{B}}_{kk}^{'\rm{H}} \mathbf{\Xi}_{k}^{-1} \bar{\mathbf{B}}'_{kk}\boldsymbol{\Gamma}_{k}\right|.
\end{equation}
Here, we see that, as the $\boldsymbol{\Gamma}_{k}$ matrices appear as multiplications from the left and right, they choose only the corresponding columns of the $\bar{\mathbf{B}}_{kk}^{'}$ matrix. Therefore, only those respective elements of $\bar{\mathbf{B}}_{kk}^{'\rm{H}} \mathbf{\Xi}_{k}^{-1} \bar{\mathbf{B}}'_{kk}$ enter the SE expression. We note that the $\bar{\mathbf{B}}'_{kk}$ matrices have the dimension $M \times LM$ and this multiplication operation selects among the effective channels between each AP $j$ and user $k$ for all $j,k$.

\subsection{Downlink Achievable SE with Perfect CSI at the Receiver}
We now particularize the upper bound SE expression, obtained with perfect CSI at the receiver, for separate stream transmission. 
The downlink received signal in \eqref{downlink_transmitted_signal_2} can be written as
\begin{equation} 
\begin{split}
     \mathbf{y}_{k} =   \mathbf{B}'_{kk} \boldsymbol{\Gamma}_{k}\boldsymbol{\varsigma}_{k} + \mathbf{n}''_{k},
\end{split}
\end{equation}
where $\mathbf{n}''_{k} = \sum_{i=1, i \neq k}^{K}  \mathbf{H}_{k}^{\rm{H}} \mathbf{W}_{i}\boldsymbol{\Gamma}_{i} \boldsymbol{\varsigma}_{i} +\mathbf{n}_{k}$ is the colored noise term consisting of the independent noise and interference terms. We can write Lemma \ref{second_lemma} using selection matrices as
\begin{equation} \label{Perfect_CSI_SE_opt2_variation}
    \mathrm{SE}_{k}^\mathrm{fullCSI} = \left(1 - \frac{\tau_{\mathrm{p}}}{\tau_{\mathrm{c}}}\right) \mathbb{E}\left\{  \log_{2} \left|  \mathbf{I}_{M} +  \boldsymbol{\Gamma}_{k}^{\rm{H}}\mathbf{B}_{kk}^{'\rm{H}}\Tilde{\mathbf{\Xi}}_{k}^{-1} \mathbf{B}'_{kk}\boldsymbol{\Gamma}_{k}  \right| \right\},
\end{equation}
where the covariance matrix $\Tilde{\mathbf{\Xi}}_{k}$ of $\mathbf{n}''_{k}$ is
\begin{align}
     \Tilde{\mathbf{\Xi}}_{k} 
     &=  \sum_{i=1, i \neq k}^{K}  \mathbf{B}'_{ki} \boldsymbol{\Gamma}_{i}\boldsymbol{\Gamma}_{i}^{\rm{H}} \mathbf{B}_{ki}^{'\rm{H}} +\mathbf{I}_{M}.
\end{align}
Similar to \eqref{eq: first_lemma_opt2_variation}, we see that sending separate streams results in only considering the respective selected blocks of $\mathbf{B}_{kk}^{'\rm{H}}\Tilde{\mathbf{\Xi}}_{k}^{-1} \mathbf{B}'_{kk}$.

\subsection{Downlink Achievable SE with Effective Channel Estimate at the Receiver}

Next, we give an achievable SE relation with a general combiner for the same case as in the former section, following a similar approach. The received signal in \eqref{downlink_transmitted_signal_2} after combiner can be written as
\begin{align} \label{eq:observation_equation_after_general_combiner_option 2}
      \tilde{\mathbf{y}}_{k} &=  \underbrace{\mathbf{U}_{k}^{\rm{H}}\hat{\mathbf{B}}'_{kk} }_{\triangleq \mathbf{E}'_{kk}} \boldsymbol{\Gamma}_{k} \boldsymbol{\varsigma}_{k} + \mathbf{U}_{k}^{\rm{H}}  \Big(  \tilde{\mathbf{B}}'_{kk}  \boldsymbol{\Gamma}_{k}\boldsymbol{\varsigma}_{k} + \hspace{-0.5em}\sum_{i=1, i \neq k}^{K} \hspace{-0.5em}\mathbf{B}'_{ki} \boldsymbol{\Gamma}_{i}\boldsymbol{\varsigma}_{i} +  \mathbf{n}_{k} \Big) \nonumber\\
      &=   \bar{\mathbf{E}}'_{kk} \boldsymbol{\Gamma}_{k}\boldsymbol{\varsigma}_{k} + (\mathbf{E}'_{kk} - \bar{\mathbf{E}}'_{kk})\boldsymbol{\Gamma}_{k}\boldsymbol{\varsigma}_{k}  \nonumber\\& \hspace{7em}+ \mathbf{U}_{k}^{\rm{H}}  \Big(  \tilde{\mathbf{B}}'_{kk} \boldsymbol{\Gamma}_{k}\boldsymbol{\varsigma}_{k}+ \hspace{-0.5em}\sum_{i=1, i \neq k}^{K} \hspace{-0.5em}\mathbf{B}'_{ki} \boldsymbol{\Gamma}_{i}\boldsymbol{\varsigma}_{i} +  \mathbf{n}_{k} \Big)\nonumber\\
      &=  \bar{\mathbf{E}}'_{kk} \boldsymbol{\Gamma}_{k}\boldsymbol{\varsigma}_{k} + \mathbf{n}''_{k},
\end{align}
where $\mathbf{E}'_{kk} \in \mathbb{C}^{M \times LM}$ denotes the effective channel after combining, \textcolor{black}{$\bar{\mathbf{E}}'_{kk}$ denotes the mean of it,} and $\mathbf{n}''_{k} \in \mathbb{C}^{M \times 1}$ the unknown part of the effective channel, interference, mismatch between the effective channel and its mean and noise after combining. \textcolor{black}{The combiner $\mathbf{U}_k$ is found with \eqref{eq:combiner for the effective channel} using the effective channel $\mathbf{B}'_{kk}$ instead.}

By following a UatF-like approach, we can write the achievable SE as
\begin{equation} \label{lemma_main_result_general_combiner_option 2}
    \mathrm{SE}_{k}^\mathrm{pilots} = \left(1 - \frac{2\tau_{\mathrm{p}}}{\tau_{\mathrm{c}}}\right) \log_{2} \left| \mathbf{I}_{M} + \boldsymbol{\Gamma}_{k}^{\rm{H}}\bar{\mathbf{E}}_{kk}^{'\rm{H}}\mathbf{C}_{\mathbf{n}''_{k}}^{-1}\bar{\mathbf{E}}^{'}_{kk} \boldsymbol{\Gamma}_{k} \right| , 
\end{equation}
where $\mathbf{C}_{\mathbf{n}''_{k}} \in \mathbb{C}^{M \times M}$ is the covariance matrix of the colored noise term defined in \eqref{eq:observation_equation_after_general_combiner_option 2}. This relation is a special case of Theorem \ref{theorem:mainresult_general_combiner}. As the effective channel has the $\boldsymbol{\Gamma}_{k}$ matrix inside, it again selects the respective subspaces of the channel. In all the relations above, the selection matrices inside the colored noise terms appear as $\boldsymbol{\Gamma}_{k}\boldsymbol{\Gamma}_{k}^{\rm{H}}$ which is equal to $\mathrm{diag}(\boldsymbol{\Gamma}_{1k},...,\boldsymbol{\Gamma}_{Lk})$. This relation allows only the respective interference terms to enter the colored noise covariance. We also note that the $\boldsymbol{\Gamma}_{k}$ matrices and selection matrices in general are not invertible. \textcolor{black}{For the separate stream case, each user estimates a set of effective channels to form the combiner. Since the set doesn't grow with the network size, this approach ensures scalability.}

\section{Precoding Schemes} \label{sec: Channel and Precoder Arguments for Option 2  Option 3}
In this section, we will first propose an MMSE precoder to be used for the same stream transmission and then define two types of precoders for separate stream transmissions with and without CSI-sharing. Finally, we exemplify channel conditions for which the same and separate stream transmission cases perform the same.  \textcolor{black}{We note that the precoders in this section are derived for the channel matrices rather than vectors. Using channel matrices inside the MMSE relations changes the derivation steps and assumptions.
Moreover, for the separate stream transmission, only certain columns of the channels are used for the filtering with the use of selection matrices. Derivations are based on the uplink-downlink duality principle using the side information of the respective method. Derivations are given in Appendix \ref{appendix: MMSE Precoder}.}

\subsection{MMSE Precoding for Same Stream Transmission} \label{MMSE Precoder for Option 1}
In the same stream transmission case, the APs share CSI that we can use to compute the precoder. We can obtain the MMSE precoder for user $k$ as
\begin{equation}
    \overline{\mathbf{W}}_{k}=q_{k} \left[ \sum_{i=1}^{K} q_{i} \Big(   \hat{\mathbf{H}}_{i}\hat{\mathbf{H}}_{i}^{H}   + \, \mathbf{C}_{\tilde{\mathbf{H}}_{i}} \Big) +  \mathbf{I}_{LN}  \right] ^{-1} \hat{\mathbf{H}}_{k},
\end{equation}
where $\mathbf{C}_{\tilde{\mathbf{H}}_{i}}=\mathbb{E}\left\{\tilde{\mathbf{H}}_{i} \tilde{\mathbf{H}}_{i}^{\rm{H}} \right\}$ captures the uplink channel estimation error for user $i$ and $q_{i}$ is the transmit power of user $i$ normalized by the noise power in the virtual uplink. The derivation of this precoder can be found in Appendix \ref{appendix: MMSE Precoder} by setting $\boldsymbol{\Gamma}_{ji}=\mathbf{I}_{M}$ and using CSI of all APs as side information. 
The per-AP power constraints, defined in Section \ref{sec: Downlink Data Transmission}, must be satisfied.
Hence, we normalize the precoders to user $k$ as 
\begin{equation}
\mathbf{W}_{k} = \sqrt{\rho_{k}}\frac{\overline{\mathbf{W}}_{k}}{\sqrt{ \sum_{l=1}^{L} \mathbb{E} \left\{ \Vert\overline{\mathbf{W}}_{lk}\Vert^{2}_{F} \right\}}},
\end{equation}
where $\overline{\mathbf{W}}_{lk}$ is the precoder from AP $l$ to user $k$ and $\rho_{k}$ determines how much power is allocated to this user.
In the numerical results, we consider
\begin{equation}
    \rho_{k}=\rho_{d}\frac{\sqrt{\sum_{l=1}^{L}\beta_{lk}}}{\max\limits_{l}\sum_{i=1}^{K}\sqrt{\sum_{l=1}^{L}\beta_{li}}}, 
\end{equation}
which slightly favors users with better channel conditions. \textcolor{black}{
Furthermore, we apply a simple scheduling algorithm where we iteratively evaluate the system performance with varying numbers of streams. Starting with the full $M$ streams per user, we calculate the sum SE using (20). We then progressively reduce the number of streams by removing one stream at a time from the last column of the precoder. 
We go through the users sequentially and determine if discarding a stream leads to a larger sum SE, in which case it is omitted.
After all users have been considered, the algorithm proceeds with subsequent iterations until no further streams are discarded. This technique mitigates the performance degradation caused by excessive inter-user interference, particularly in scenarios where the spatial dimensions available $LN$ are insufficient to support all potential streams $2KM$.}


\subsection{MMSE Precoding with CSI-Sharing and Separate Streams} \label{MMSE-like Precoder for Option 3}
For the separate stream transmission with  CSI sharing, the MMSE precoder used at AP $l$ for user $k$ takes the form
\begin{equation} \label{eq: MMSE Precoder for Option 3}
    \overline{\mathbf{W}}'_{lk}=q_{k} \left[\sum_{j=1}^{L}\sum_{i=1}^{K} q_{i}   \Big(\hat{\mathbf{H}}_{ji}\boldsymbol{\Gamma}_{ji}\hat{\mathbf{H}}_{ji}^\mathrm{H}  +\mathbf{C_{\tilde{\mathbf{H}}_{ji}\boldsymbol{\Gamma}_{ji}}} \Big) +   \mathbf{I}_{N}\right]^{-1}\hspace{-0.5em} \hat{\mathbf{H}}_{lk},
\end{equation}
where $\mathbf{C_{\tilde{\mathbf{H}}_{ji}\boldsymbol{\Gamma}_{ji}}}=\mathbb{E}\left\{ \tilde{\mathbf{H}}_{ji} \boldsymbol{\Gamma}_{ji} \tilde{\mathbf{H}}_{ji}^\mathrm{H}\right\}$ is the covariance of used column vectors of the channel estimation error between AP $j$ and user $i$.
The precoders are normalized as 
\begin{equation}
\mathbf{W}'_{lk} = \sqrt{\rho_{lk}}\frac{\overline{\mathbf{W}}'_{lk}}{\sqrt{ \mathbb{E} \left\{ \Vert\overline{\mathbf{W}}'_{lk}\Vert^{2}_{F} \right\}}},
\end{equation}
where the coefficients $\rho_{lk}$ must be selected to satisfy the per-AP power constraints.

One can use different ways to choose the $\boldsymbol{\Gamma}_{lk}$ matrices and power control coefficients. In the numerical results section, we use
the large-scale fading coefficients $\beta_{lk}=\frac{\mathrm{tr}(\mathbf{R}_{lk})}{N}$ for this purpose. The $M$ APs with the highest $\beta_{lk}$ values transmit one stream each to user $k$, and we form the $\boldsymbol{\Gamma}_{lk}$ matrices accordingly. Moreover, we choose the power control coefficients as $\rho_{lk}=\rho_{d}\frac{\mathrm{rank}(\boldsymbol{\Gamma}_{lk})\sqrt{\beta_{lk}}}{\sum_{i=1}^{K}\mathrm{rank}(\boldsymbol{\Gamma}_{li})\sqrt{\beta_{li}}}$. This normalization method also prioritizes users with more favorable channel conditions. Furthermore, since the number of streams transmitted from AP $l$ to user $k$ is equal to $\mathrm{rank}(\boldsymbol{\Gamma}_{lk})$, the power allocation to users is proportional to this value. The per-AP power constraints discussed in Section \ref{sec: Downlink Data Transmission} are satisfied with equality at each AP, allowing them to transmit at maximum power without being subject to a joint power constraint. This is generally not the case in the same stream transmission case, where we are only guaranteed that one AP uses its maximum power.
The normalized precoder from AP $l$ to user $k$ is given by $\mathbf{W}'_{lk} = \rho_{lk}\overline{\mathbf{W}}'_{lk}$.


The derivation of this precoder can be found in Appendix \ref{appendix: MMSE Precoder}. The precoder $\mathbf{W}_{k}$ obtained with this transmission method is block diagonal, as the precoders $\mathbf{W}_{lk}=\mathbf{W}'_{lk}\boldsymbol{\Gamma}_{lk}$ found here are stacked to form a block diagonal matrix. The $\boldsymbol{\Gamma}_{ji}$ matrices select the specific columns of each channel to which transmission is allowed and suppress interference in those directions. Moreover, we only consider the channel estimation errors in those directions. 

\subsection{MMSE Precoding with no CSI-Sharing and Separate Streams}\label{MMSE-like Precoder for Option 2}
For the separate stream transmission without CSI sharing, each AP must compute its precoder solely based on the locally available CSI. The interference caused by other APs is only represented statistically. For this case, the MMSE precoder takes the form
\begin{align} \label{eq: MMSE Precoder for Option 2} 
    \overline{\mathbf{W}}'_{lk} =q_{k} \Bigg[&\sum_{i=1}^{K} q_{i}  \Big(\hat{\mathbf{H}}_{li}\boldsymbol{\Gamma}_{li}\hat{\mathbf{H}}_{li}^\mathrm{H}  +\mathbf{C_{\tilde{\mathbf{H}}_{li}\boldsymbol{\Gamma}_{li}}} \Big) \nonumber \\ &+ \sum_{i=1}^{K}\sum_{j=1, j\neq l}^{L} q_{i}  \mathbb{E}\left\{ \mathbf{H}_{ji} \boldsymbol{\Gamma}_{ji}\mathbf{H}_{ji}^\mathrm{H} \right\} +   \mathbf{I}_{N}\Bigg]^{-1} \hat{\mathbf{H}}_{lk}.
\end{align}
The derivation of this precoder can be found in Appendix \ref{appendix: MMSE Precoder}. We note that the precoder $\mathbf{W}_{k}$ for this transmission method is again block diagonal. It is structured and scaled as explained in the last subsection. 
As in that subsection, the $\boldsymbol{\Gamma}_{ji}$ matrices choose the subspaces of interference and channel estimation errors to be suppressed. Since there is no CSI related to other APs, the statistics of the respective subspaces are used to suppress the interference caused by other APs to user $k$.

\textcolor{black}{The MMSE precoders introduced in this section minimize the MSE of the estimated data symbols conditioned on different types of side information. For the same stream transmission, the MSE is minimized with global CSI at all APs. For the separate stream transmission with CSI sharing, the MSE is minimized given the shared CSI across APs. For the case without CSI sharing, the MSE is minimized at each AP based only on its locally available CSI. These formulations adapt the MSE minimization to the available information, ensuring optimality under the specific operational constraints of each method.
}

\subsection{A Channel Condition for Equivalent Performance of Same Stream and Separate Stream Transmissions} \label{SVD Precoder}

In general, the same stream and separate stream transmission methods would yield different precoders. However, certain channel structures can make these two transmission schemes perform identically. We note that separate stream transmission has block diagonal precoders by definition. Therefore, if the channel matrix has such a structure from the transmitter's perspective, the same stream transmission will become equivalent to the separate stream transmission.
This occurs when the channels between the serving APs and user $k$ span orthogonal spatial dimensions. 
This can be observed by examining the right singular vectors of the channel. 
Let $\mathbf{H}_{jk}^{\mathrm{H}} = \mathbf{U}_{jk} \boldsymbol{\Sigma}_{jk} \mathbf{V}_{jk}^\mathrm{H}$ denote the SVD of each AP's channel to user $k$, then we have orthogonal spatial transmit dimensions if the SVD of user $k$'s complete channel matrix can be written as
\begin{align} \label{eq: SVD of block diagonal channel}
   \setlength{\fboxsep}{0pt}
\mathbf{H}_{k}^\mathrm{H} &= \mathbf{U}_{k} \boldsymbol{\Sigma}_{k} \mathbf{V}_{k}^\mathrm{H},
\end{align}
where $\mathbf{U}_{k}=[\mathbf{U}_{1k}, \ldots,   \mathbf{U}_{L_{k} k}]$, $\boldsymbol{\Sigma}_{k}=\rm{diag}(\boldsymbol{\Sigma}_{1k},\ldots,\boldsymbol{\Sigma}_{L_{k}k})$ and $\mathbf{V}_{k}=\rm{diag}(\mathbf{V}_{1k},\ldots,\mathbf{V}_{L_{k}k})$. We assume $L_{k}=L= M$ is the total number of APs serving user $k$ for this example case. For a channel with this structure, the matrix $\mathbf{V}_{k}^\mathrm{H}$ with the right singular vectors is block diagonal, with each block associated with one AP. 
This implies that signals transmitted from two different APs cannot be received in the same subspace and therefore cannot add constructively.
Consequently, when the channel has the block-diagonal structure in \eqref{eq: SVD of block diagonal channel}, the optimal same stream transmission precoder automatically becomes block diagonal and coincides with the optimal separate stream transmission precoder.
A channel with this structure can appear exactly or approximately when the user sees the APs in very different angular directions, for example, in a connected stadium where a few APs are equally spaced around the perimeter of the stadium.

\textcolor{black}{\section{Computational Costs}
We will now quantify the computational complexity per coherence block when using the proposed schemes.
The matrices used in uplink channel estimation depend on the statistics of the channels.
Processing the received signal during the pilot transmission requires $N\tau_p$ complex multiplication per pilot sequence, and the MMSE channel estimation process requires $(NM)^2$ complex multiplications. Therefore, the complexity for the uplink channel estimation is $K((NM)^2+N\tau_p)$ complex multiplications per coherence block.
}

\textcolor{black}{The complexity of computing the precoder matrices is dominated by the $LN\times LN$ matrix inversions. The number of complex multiplications per coherence block is
\begin{align*}
\frac{(NL)^2 + NL}{2} MK + (NL)^2 M + \frac{(NL)^3 - NL}{3}.
\end{align*}
From this relation, we can see that the complexity increases linearly with $\tau_p$ as it is proportional to $M$}

\begin{table}[t]
\centering
\textcolor{black}{ 
\centering
\caption{comparison of fronthaul communication load (total number of complex scalars) between the same-stream and the separate-stream transmission schemes.}
\centering
\begin{tabular}{lcc}
\hline
Scheme & Pilot Signaling & Data Signaling \\
\hline
Same-stream & $\tau_p N M$ & $(\tau_c - \tau_p) N $ \\
Separate-stream & $\tau_p N M$ & $ (\tau_c - \tau_p) N$ \\
\hline
\end{tabular}
\label{tab:fronthaul_comparison}
}
\end{table}

\textcolor{black}{The fronthaul signaling related to sending the received pilot signals from each AP to the CPU is $\tau_pN$ complex scalars for the same stream transmission case. The fronthaul signaling for the precoded downlink data signals necessitates sending $N\times 1$ vectors comprising the multiplication of the precoder matrix and the data vector. Therefore, the fronthaul load for the data signals is $(\tau_c-\tau_p)N$ complex scalars. This value is irrespective of the choice of the precoding scheme.}

\textcolor{black}{For the separate stream case, the fronthaul signaling for the uplink channel estimation and sending the data is the same. This happens as we multiply the precoder matrix by the data vector.
A comparison of these can be seen in Table \ref{tab:fronthaul_comparison}.}

\section{Numerical Results} \label{Numerical_Results}

In this section, we provide numerical results where we will compare the downlink SEs achieved with different transmission methods and different numbers of antennas and APs. We use the same simulation setup as in \cite{demir2021foundations}.
Hence, we assume that the AP and user locations are uniformly distributed in a $1 \times 1 \; \text{km}^2$ area.  We use $\tau_{\mathrm{p}}$ mutually orthogonal pilot sequences that are shared among the users.
Pilot assignment is done as explained in \cite[Ch. 4]{demir2021foundations} and each pilot sequence is used by $K'$ users, which is an integer value chosen as a ratio of $K$.
Therefore, the pilot length is $\tau_{\mathrm{p}} = K' M$. The same pilot length and assignment are used for uplink and downlink channel estimation. 
  \textcolor{black}{We adopt the pilot assignment scheme from \cite{demir2021foundations}, as simple allocation methods are remarkably effective in cell-free systems. This efficacy stems from the pronounced pathloss variations inherent to the architecture, which create a natural spatial separation among users and thereby intrinsically limit pilot contamination. Furthermore, our use of MMSE channel estimation provides a powerful, built-in mechanism for decontamination that uses the channel covariance matrices. As established in \cite{8094949}, the MMSE estimator statistically separates interfering users by exploiting their distinct spatial channel statistics. While our approach already exploits these statistics for decontamination, one could alternatively use the covariance matrices $\mathbf{R}_{lk}$ for pilot design in cell-free massive MIMO systems. For instance, a correlation-aware pilot assignment method could reduce pilot overhead or contamination by assigning pilots based on the similarity of users' spatial correlation structures, ensuring that users with overlapping channel subspaces receive minimally interfering pilots. This mitigates inter-user interference during estimation without necessitating additional pilot resources. Such approaches are explored in \cite{9160993,6415397}. Moreover, subspace-based training methods could reduce pilot overhead or contamination by projecting pilots onto the dominant eigenspaces of $\mathbf{R}_{lk}$, allowing for training sequences that target the low-rank channel structure rather than the full dimension, which is particularly effective for multi-antenna users to lower the required pilot length while improving estimation accuracy in the presence of spatial correlation. These techniques are demonstrated in \cite{9110802,11044434}. We do not incorporate these additional techniques in this study in order to maintain a manageable system complexity and, more importantly, to clearly isolate the performance gains attributable to the core contributions of this paper.}
For downlink transmission, the power control coefficients explained in Section \ref{sec: Channel and Precoder Arguments for Option 2  Option 3} are used.

We will primarily compare the following three proposed transmission methods in this section:
\begin{itemize} 
\item \textbf{Method 1:} Same stream transmission with the precoder given in Section \ref{MMSE Precoder for Option 1}.
\item \textbf{Method 2:} Separate stream transmission without CSI sharing, with the precoder explained in Section \ref{MMSE-like Precoder for Option 2}.
\item \textbf{Method 3:} Separate stream transmission with CSI sharing, with the precoder explained in Section \ref{MMSE-like Precoder for Option 3}. 
\end{itemize}
Table \ref{table_methods} summarizes the main differences between these transmission methods, which are related to what information must be shared between the APs when implementing them.

\begin{table}
\centering
\caption{Comparison of Options}
\label{table_methods}
\begin{tabular}{ccccc} 
\toprule
                  & Method 1                  & Method 2                  & Method 3                   \\ 
\midrule
CSI Sharing       & \checkmark               & $\boldsymbol{\times}$                  & \checkmark                \\ 
Data Sharing     &         \checkmark               &      $\boldsymbol{\times}$          &           $\boldsymbol{\times}$              &  \\ 
\bottomrule
\end{tabular}
\end{table}
Moreover, the total number of streams $M$ is less than the number of APs $L$. We assume that $L=20$, $N=4$, $M=2$, and $K=5$ unless stated otherwise. Furthermore, we assume that $\tau_{p}=KM/2$ and the correlated Rayleigh fading channels are modeled as in \cite[Sec. 5.2]{demir2021foundations} with an angular standard deviation (ASD) of $15^{\circ}$. We consider the bounds we derived with the downlink channel estimate for the results, unless otherwise stated. \textcolor{black}{The total number of AP antennas bounds the multiplexing gain. With $M$ antennas per user and $K$ users, our configuration supports up to $MK$ data streams, which is the metric to compare with previous works. Our setup can be viewed as a subset of a larger cell-free massive MIMO network operating on a specific time-frequency resource block. In practice, resource allocation and scheduling partition a larger number of APs and UEs into smaller groups, ensuring that not all users are active simultaneously.}

\subsection{Differences Between the Proposed Methods}

\begin{figure}[t!]
\centering
  \includegraphics[width=\linewidth]{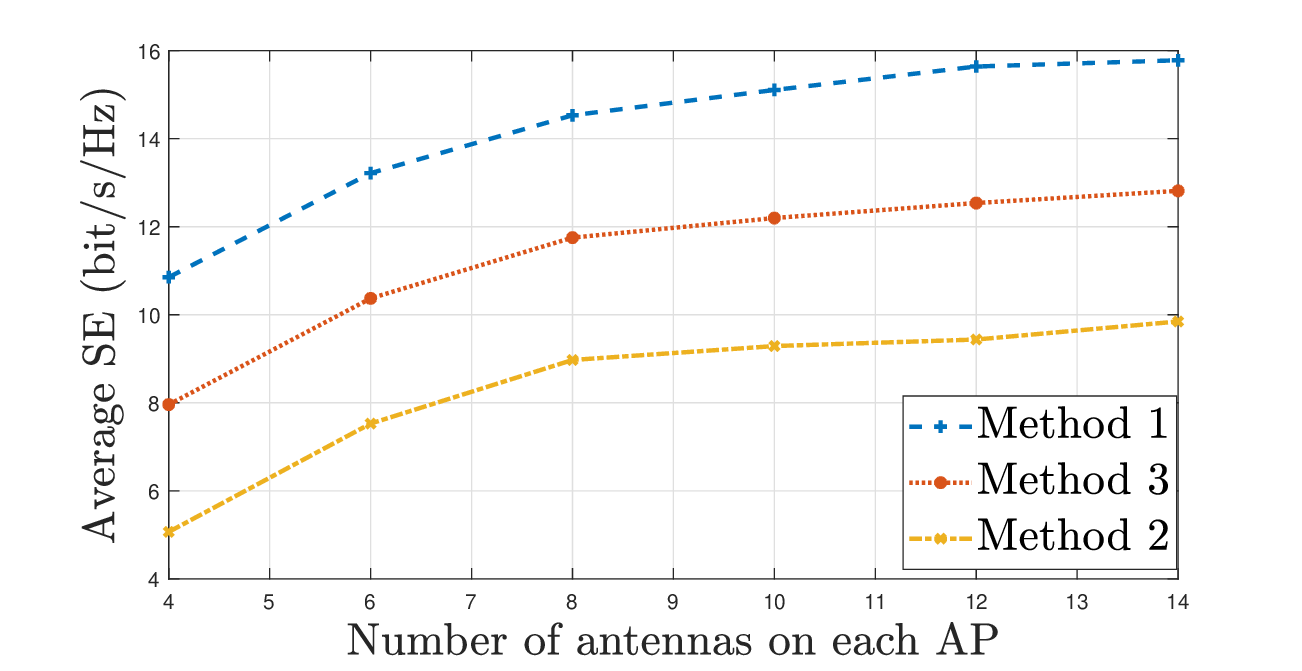} \vspace{-6mm}
  \caption{Comparison of the average per-user SEs achieved with Methods 1-3 as a function of the number of AP antennas and with $M=2$ user antennas.} \vspace{-3mm}
  \label{fig:figureB}
\end{figure}

We first investigate the impact of the choice of transmission methods on the SEs.
Fig.~\ref{fig:figureB} shows the average SE of a randomly located user as a function of the number of antennas per AP \textcolor{black}{with the MMSE precoders}.
From this figure, we can see that Method 1 performs better than the other methods. This outcome is anticipated as Method 1 builds on sharing CSI among all APs so that the centralized MMSE precoder can be utilized to suppress interference in a coordinated manner. Method 3 provides slightly lower SEs than Method 1, as the lack of data sharing forces the precoding matrix to be block diagonal. However, Method 3 can still suppress inter-AP interference since CSI is shared to enable that. One clear benefit of Method 3 is that there is no need for phase synchronization between the APs, while for Method 1, synchronization is crucial for the APs to jointly beamform signals that add constructively at the intended user and destructively at other users. Finally, Method 2 performs worse than Method 3 as it lacks CSI sharing between APs, but it can only suppress inter-AP interference based on its statistics.

\subsection{Impact of Spatial Correlation}

\begin{figure}[t!]
\centering
  \includegraphics[width=\linewidth]{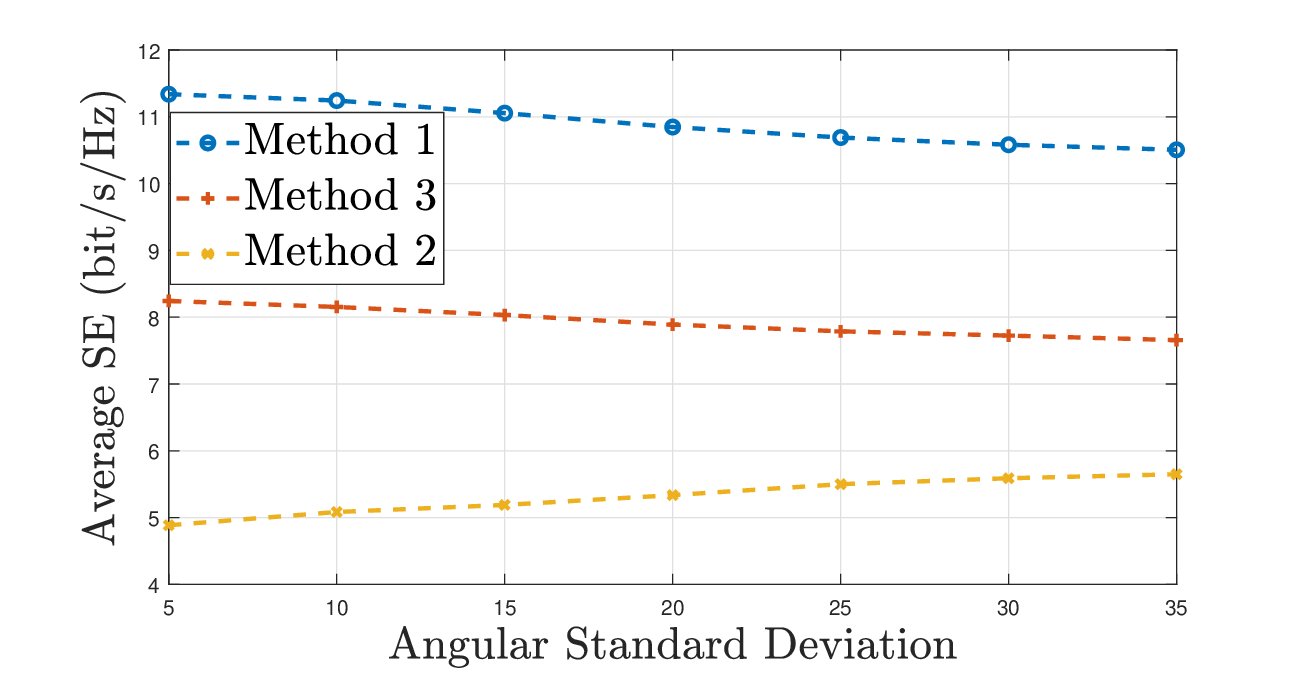} \vspace{-6mm}
  \caption{Comparison of the average per-user SEs achieved with Methods 1-3 as a function of the angular standard deviation and with $N=4$ and $M=2$.} \vspace{-2mm}
  \label{fig:figureaa}
\end{figure}

Next, we investigate the impact spatial correlation has on the SEs. Spatial correlation is modeled similarly to that in \cite[Sec. 5.2]{demir2021foundations} assuming that both AP and users are equipped with ULAs. Moreover, the spatial correlation matrix elements are computed between each AP and user antennas as explained there. The ASD variation changes the correlation from the AP's viewpoint, thereby modeling how widely the scattering objects are distributed around the user. \textcolor{black}{Fig.~\ref{fig:figureaa}} shows the average SE as a function of the ASD, where a small value means high spatial correlation and vice versa.
The figure shows that when the ASD is low, the same stream transmission and separate stream transmission with CSI sharing work particularly well. The performance then reduces as the ASD increases.
However, the separate stream transmission without CSI sharing (Method 2) performs better when the ASD is higher. 
Hence, spatial correlation can either enhance or degrade the system's performance depending on the method used.

A high spatial correlation causes a large variation in the singular values of the channel matrix. 
Moreover, it reduces the degree of channel hardening. The channel hardening is maximized when the channel coefficients at the different antennas are uncorrelated and equally distributed, leading to the maximum diversity and the singular values being roughly the same. However, high spatial correlation can help an AP to separate users from each other. Moreover, spatial correlation can also improve the quality of channel estimation. 
Here, we observe the combined impact of all these points. We will give further explanations in the subsequent subsections. 

\subsection{Impact of the Number of Users}

\begin{figure}[t!]
\centering
  \includegraphics[width=\linewidth]{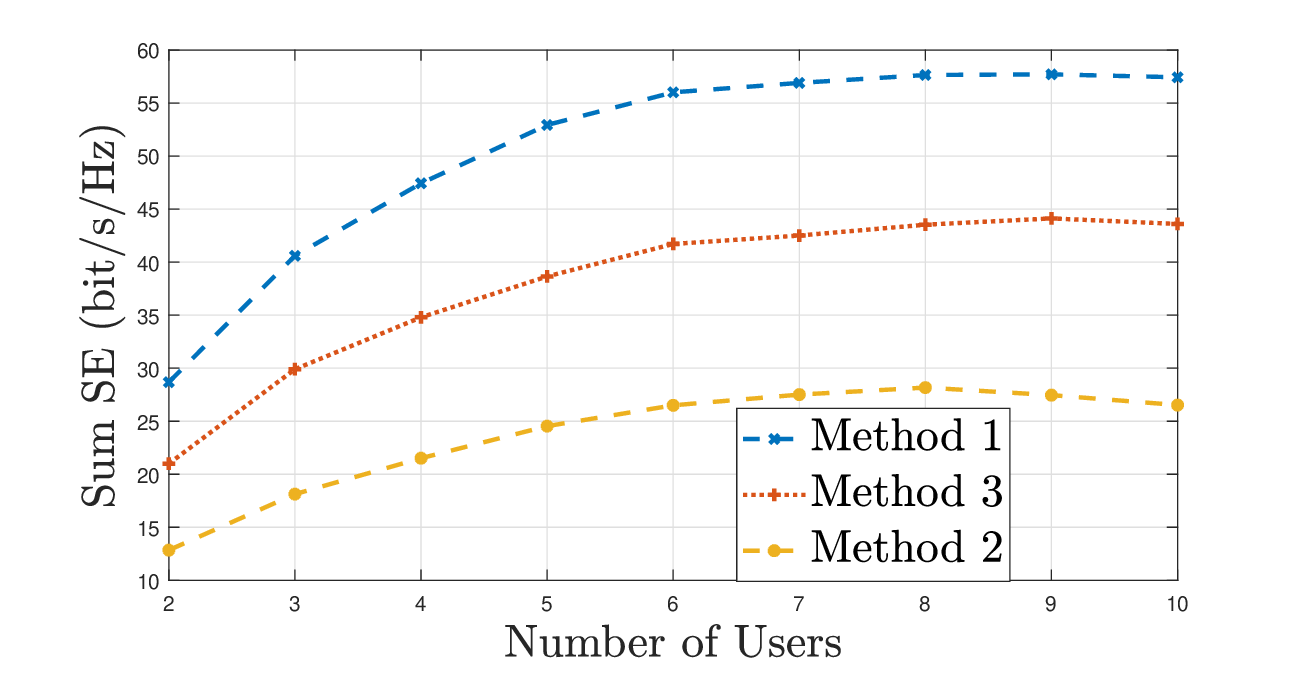} \vspace{-6mm}
  \caption{Achievable sum SE as a function of the number of users with $N=4$ and $M=2$.} \vspace{-3mm}
  \label{fig:figureA}
\end{figure}

We will now investigate the impact that the number of users has on the SEs. Fig.~\ref{fig:figureA} shows the sum SE when $K$ varies from $2$ to $10$. We assume that, as the number of users increases, the number of pilots also increases proportionally.
We notice that the sum SE increases as $K$ increases, but the curves grow slower than linear, which indicates that the
per-user SE decreases. Eventually, the sum SE reaches a plateau and then begins to decay, which is particularly visible for Method 3.
This is an expected result as the amount of interference increases significantly as we transmit more data streams. The methods with CSI sharing are also better at handling the increasing interference.

\subsection{Impact of the Number of Antennas per User}
In this subsection, we investigate the impact of the number of antennas on users with Method 1.

\begin{figure}[t!]
\centering
  \includegraphics[width=\linewidth]{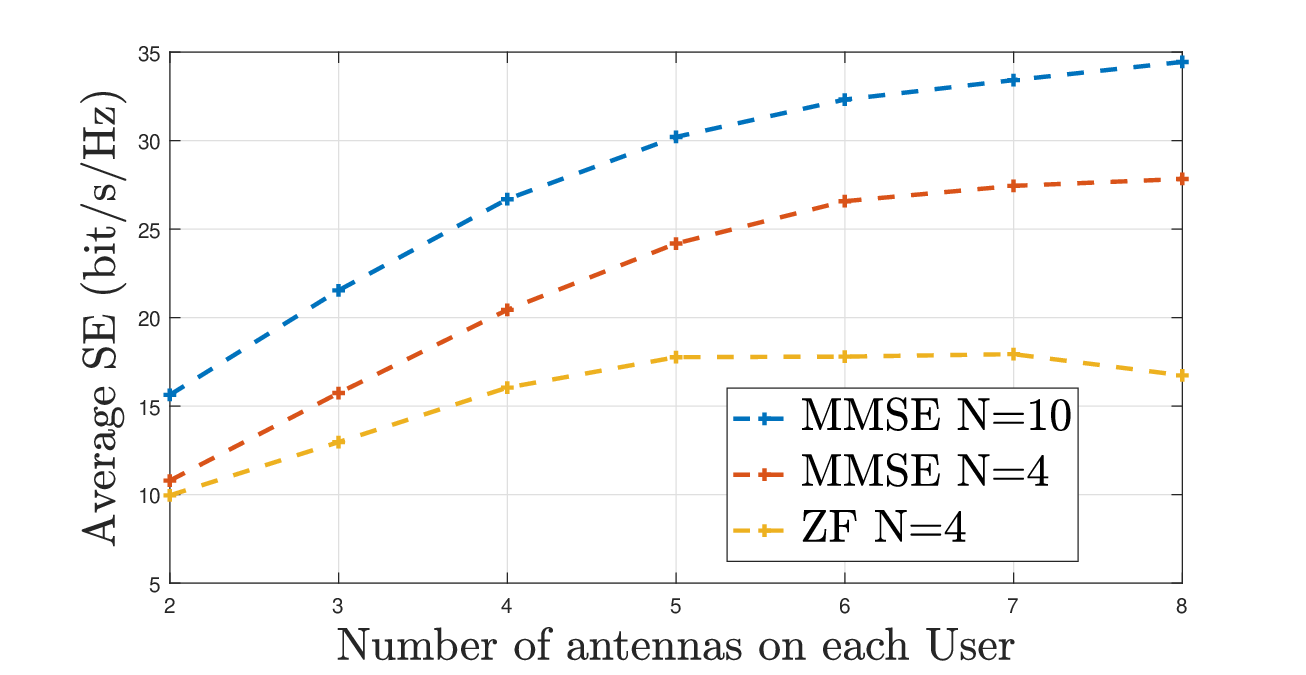} \vspace{-6mm}
  \caption{Achievable per user SE with ZF and MMSE combiners as a function of the number of user antennas.} \vspace{-3mm}
  \label{fig:figureC}
\end{figure}

Fig.~\ref{fig:figureC} shows the average SE when using either the MMSE or ZF combiner. We can clearly see that the MMSE combiner yields better performance than the ZF combiner. Hence, although the ZF combiner diagonalizes the effective channel and leads to the simpler SE expression in Corollary~\ref{th:mainresult}, it overemphasizes the need for canceling interference.
The MMSE combiner finds a better balance between suppressing and noise, taking also the estimation errors into account.

\subsection{Impact of the Coherence Block Length}

\begin{figure}[t!]
\centering
  \includegraphics[width=\linewidth]{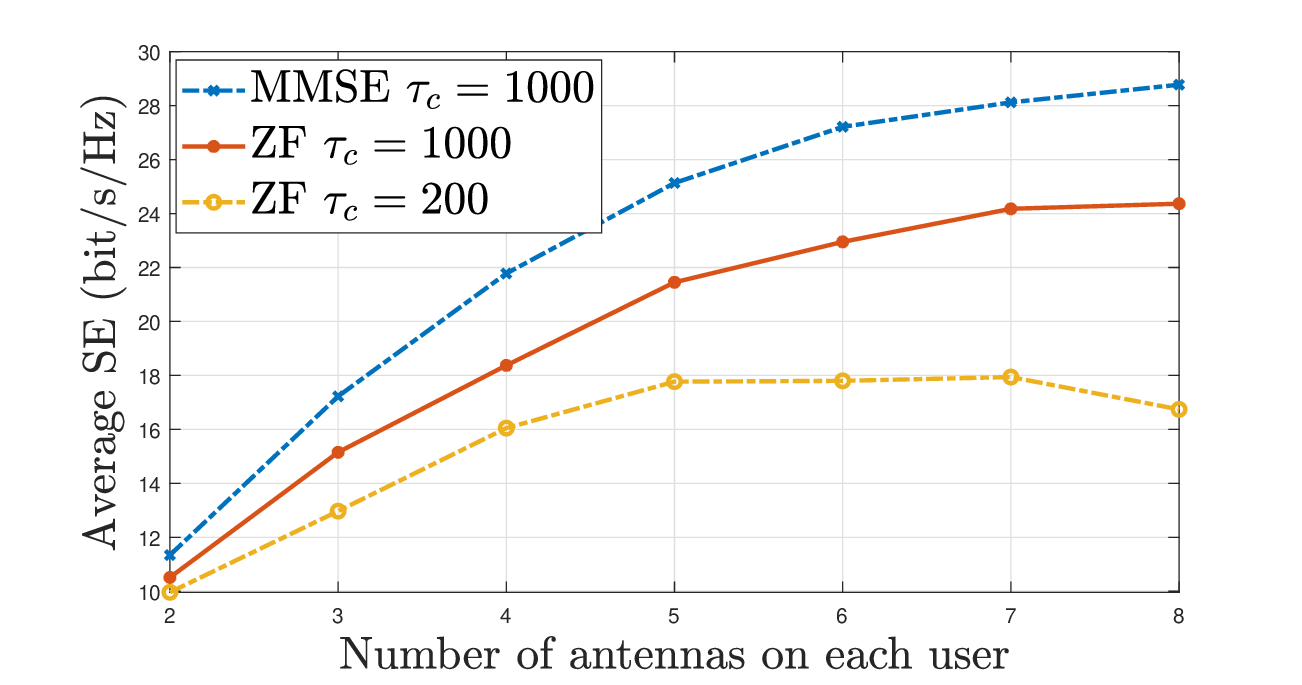} \vspace{-6mm}
  \caption{Achievable per user SE with ZF and MMSE combiners as a function of the number of user antennas.} \vspace{-3mm}
  \label{fig:figureD}
\end{figure}

In Fig.~\ref{fig:figureD}, we look at the impact of the coherence block length on the SE with the ZF combiner.
We vary the number of user antennas as in Fig.~\ref{fig:figureC} and will revisit the issue mentioned in the former subsection. 
We note that the number of pilots increases linearly with the number of user antennas.
Having more antennas should not degrade the SE, but it does when using ZF with $\tau_{\mathrm{c}}=200$. \textcolor{black}{The reason is that the SE during data transmission increases very little compared to the reduction in the pre-log factor caused by the growing pilot overhead. However, if $\tau_{\mathrm{c}}$ were higher, the extra pilot overhead would be less impactful, relatively speaking.
We can see that when $\tau_{\mathrm{c}}=1000$, the SE increases with $M$ for the considered range of antennas. Hence, the benefit of having more user antennas is particularly large in low-mobility cases. From Fig.~5 and Fig.~6, we see that the MMSE combiner is a better option than the ZF combiner. We need  $\tau_{\mathrm{c}}$ to be sufficiently larger if the ZF combiner is to be used.}

\subsection{Comparison of the Capacity Bounds}

In this section, we compare the different capacity bounds from Section~\ref{sec: SE Results for Option 1} and comment on how the amount of channel hardening depends on the number of antennas per user. 
In order to do so, we will compare our proposed SE expression with downlink pilots in Theorem \ref{theorem:mainresult_general_combiner}
with the SE results obtained without CSI in Lemma \ref{first_lemma} (i.e., the hardening bound) and with perfect CSI in Lemma \ref{second_lemma}. 

\begin{figure}[t!]
\centering
  \includegraphics[width=\linewidth]{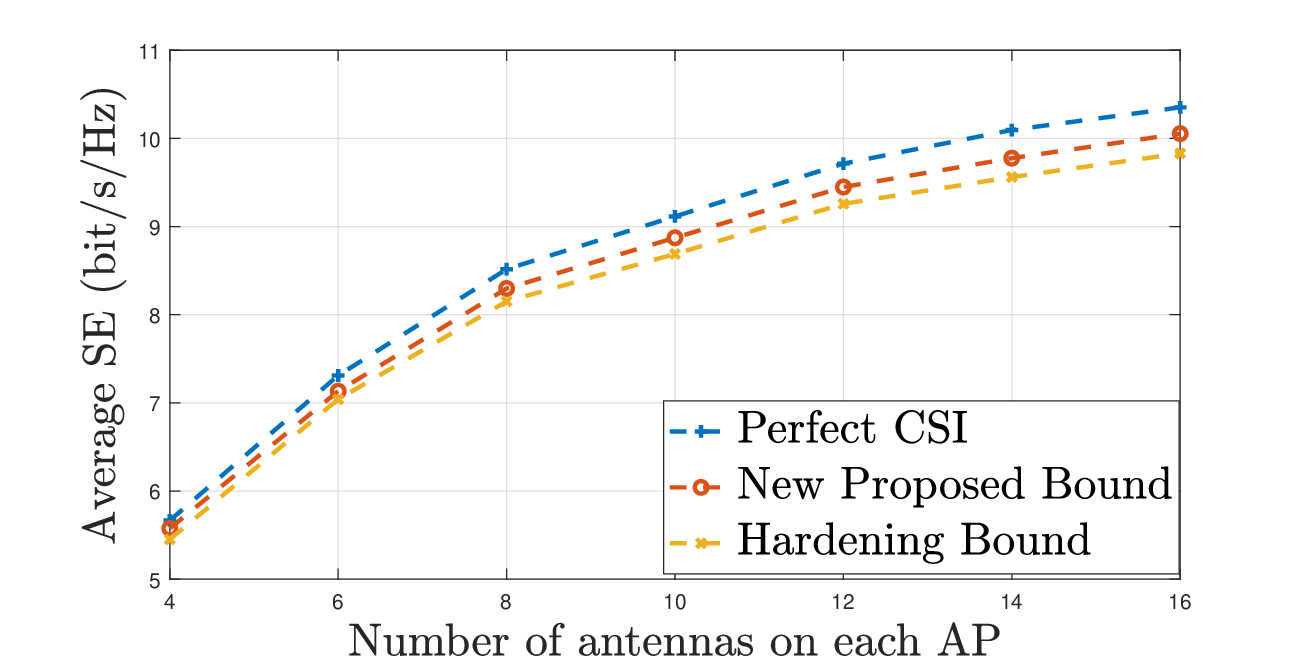} \vspace{-6mm}
  \caption{Comparison of Perfect CSI and hardening bounds for $M=1$.} \vspace{-3mm}
  \label{fig:FigureA_last}
\end{figure}

Fig.~\ref{fig:FigureA_last} shows the average SE obtained with Method 1 for $M=1$ antenna per user \textcolor{black}{with i.i.d. Rayleigh fading}. This is the most common setup in previous works, including the textbook \cite{demir2021foundations}. There is little difference between the curves since channel hardening occurs in this case, so one can omit downlink pilots and use the hardening bound without much performance loss. 

\begin{figure}[t!]
\centering
  \includegraphics[width=\linewidth]{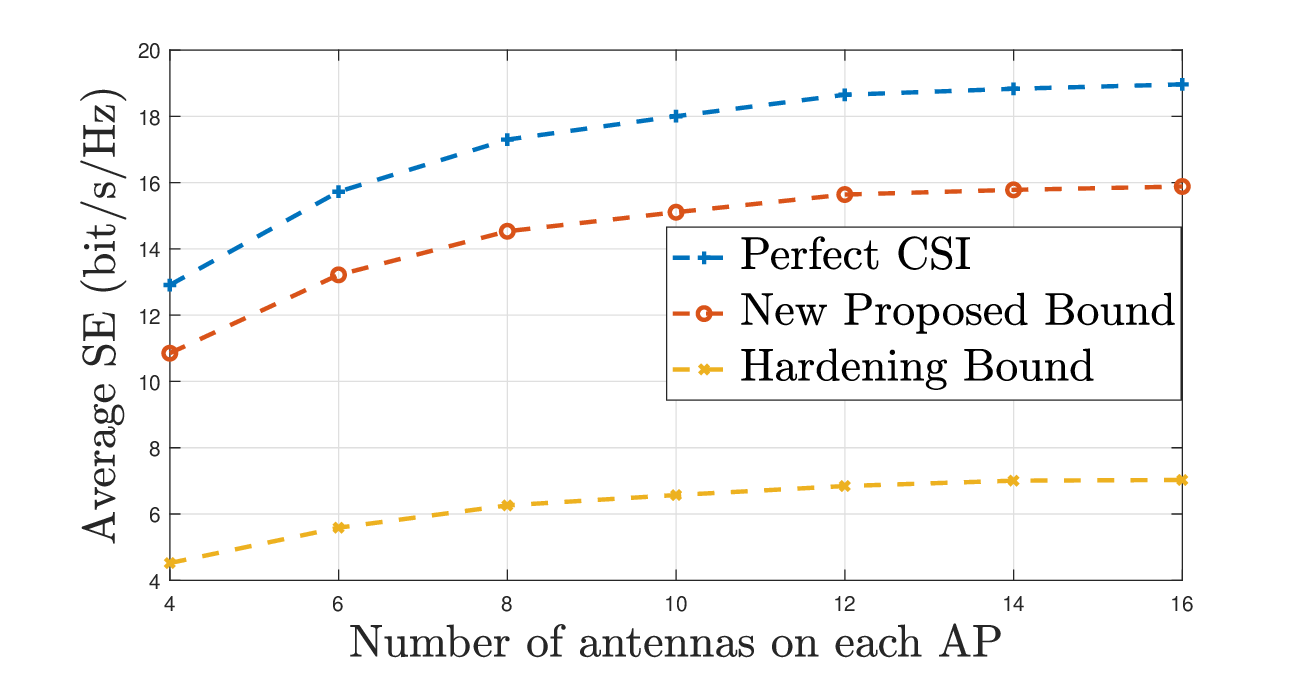} \vspace{-6mm}
  \caption[Caption for LOF]{Achievable per user SE as a function of the number of AP antennas with $M=2$.\footnotemark} \vspace{-3mm}
  \label{fig:Figure1_last}
\end{figure}

Fig.~\ref{fig:Figure1_last} considers the same setup but with $M=2$. As discussed earlier in the paper, we cannot rely on channel hardening because only the diagonal entries of the effective channel harden, while the inter-stream interference in the off-diagonal entries does not vanish. Consequently, there is a huge gap between the perfect CSI and no CSI cases for $M>1$. However, using estimates of the downlink effective channels, as proposed in Theorem \ref{theorem:mainresult_general_combiner} leads to vastly higher performance than in the case of no CSI. The difference between the perfect CSI and the proposed method is due to two primary factors. One is the ability of the combiner filter to suppress interference and mitigate estimation error, which is inherently dependent on the effective channel estimates. Secondly, it is affected by the increased channel estimation overhead. \textcolor{black}{Fig. 9 considers the case with the single AP and $N=80$ antennas, where the total number of antennas matches the cell-free massive MIMO case. We observe that the gap between the proposed method and the hardening bound remains significant in this case. Therefore, we also conclude that the gap is not due to the correlation or distribution of the APs, but rather a result of the multiple antennas on the users. These simulation results are in line with the analytical results given in Section~\ref{sec: SE Results for Option 1}}


In \cite{demir2021foundations}, it is briefly mentioned that multi-antenna users can be treated as multiple single-antenna users placed close to each other.  Here, if we assume that each serving AP transmits a separate stream to each antenna of user $k$, our model aligns with that assumption. In Fig.~\ref{fig:FigureE}, we compare the average SE achieved by this sub-optimal method with our proposed Method 3. The results are shown as a function of the number of antennas on each AP, and we consider both the proposed bound and the hardening bound.
We observe that the hardening bound is close to the proposed bound when using the sub-optimal method. 
Therefore, the system can operate without downlink pilots with this transmission method but rely on channel hardening, leading to less channel estimation overhead. However, we see that the SEs achieved by this sub-optimal transmission method are significantly lower than those of Methods 1-3. This is due to the loss of beamforming gain and increased inter-stream interference. These results tell us that we should not treat a multi-antenna user as multiple single-antenna users, and we need to do pilot-based downlink channel estimation in order to have high SE with multiple antenna users. We also observe that, although the structure of the precoder is block diagonal for Method 3, the hardening bound isn't tight. We omitted the SE results for Method 2 as these are similar to Method 3.

\footnotetext{\textcolor{black}{If one would double the pilot length for the perfect CSI case to model that we obtain perfect channel estimates by downlink pilots, the curve would drop a few percent.}}

\begin{figure}[t!]
\centering
  \includegraphics[width=\linewidth]{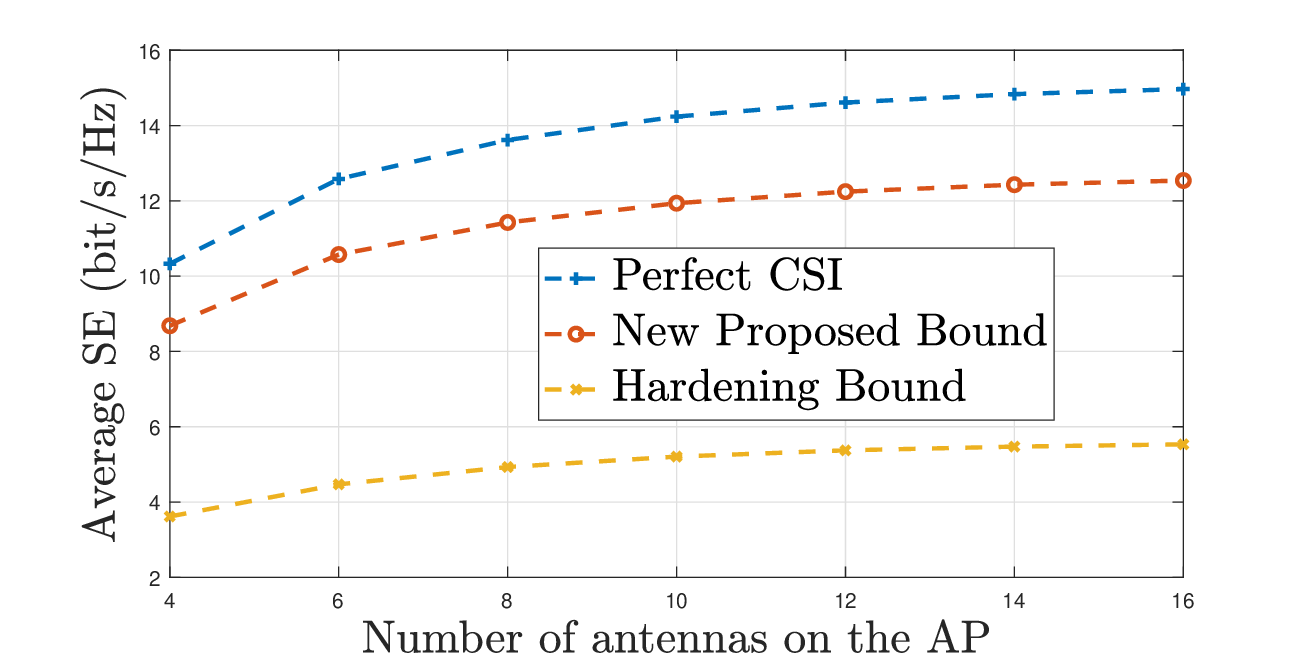} \vspace{-6mm}
  \caption{Achievable per user SE as a function of the number of AP antennas with $M=2$ and a single AP.} \vspace{-3mm}
  \label{fig:Figure1_last_mMIMO}
\end{figure}

\begin{figure}[t!]
\centering
  \includegraphics[width=\linewidth]{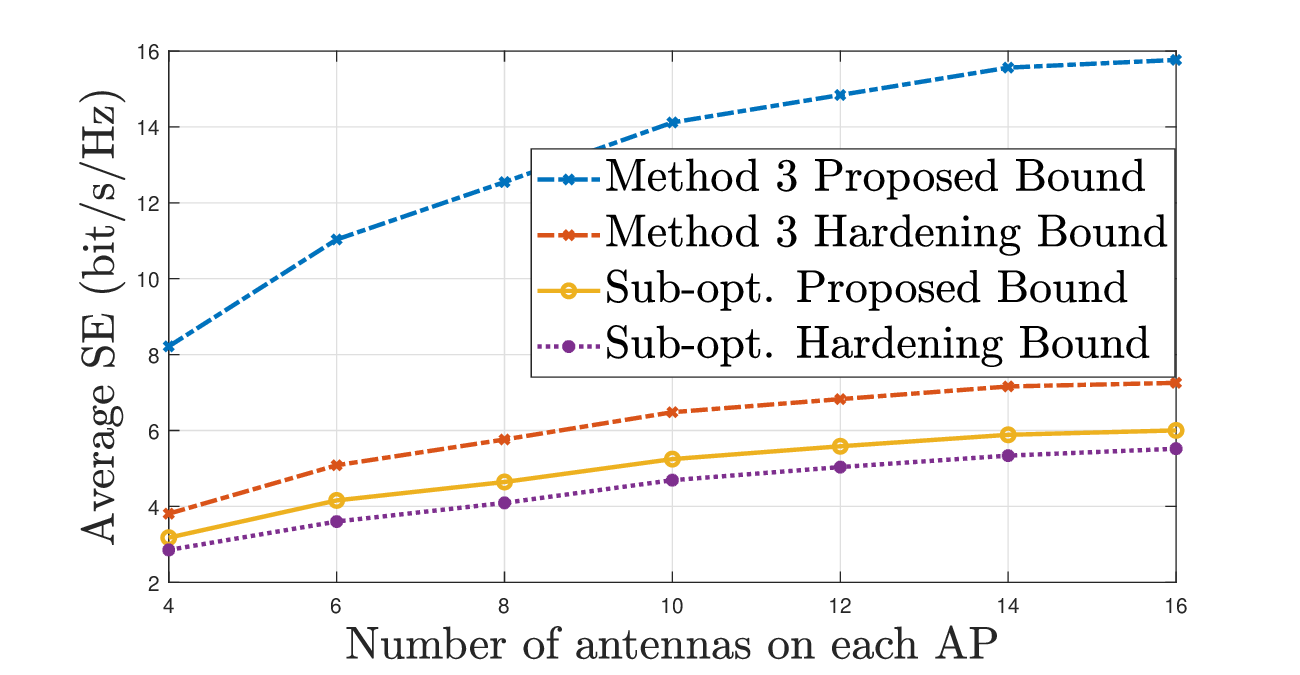} \vspace{-6mm}
  \caption{Achievable per user SE as a function of the number of AP antennas with $M=2$.} \vspace{-3mm}
  \label{fig:FigureE}
\end{figure}

\section{Conclusions} \label{sec: Conclusions}
\textcolor{black}{
In this work, we present a novel analysis of cell-free massive MIMO systems with multi-antenna users, where each user can receive multiple spatial streams. We analyzed the transmission of these streams from either multiple APs, which requires data sharing, or from separate APs, and compared CSI-sharing and no CSI-sharing in the separate stream case. MMSE precoders were derived for both transmission schemes, and stream assignment strategies were proposed accordingly. Our results demonstrate that the conventional hardening bound, effective for single-antenna users, is inadequate for multi-antenna configurations; hence, we introduced a novel pilot-based downlink LMMSE channel estimation scheme along with its corresponding capacity bound. 
Through simulations, we demonstrated how this new approach can significantly improve the SEs and deliver performance similar to the perfect CSI case. We compared the three transmission methods and demonstrated that the same stream method outperforms both other methods. However, the separate stream transmission with the CSI-sharing method performs close to the same stream transmission, and it has less fronthaul requirement. Therefore, it is a feasible option.
We have also showcased the impact of varying numbers of users, APs, ASD, and user and AP antennas. We have seen that even with the high number of users and user antennas, we can achieve good performance despite the increased amount of interference. Moreover, we have seen that spatial correlation has a significant impact on performance as the interference here is both among user antennas and users.
}



\appendix
\section{Appendix}

\subsection{Proof of Theorem 1} \label{Appendix: Proof of main result general combiner}

The capacity is defined as
\begin{equation} \label{capacity_definition_lemma3}
    C = \max_{p(\boldsymbol{\varsigma}_{k}|\bar{\mathbf{E}}_{kk})} I(\boldsymbol{\varsigma}_{k};\mathbf{y}_{k}, \bar{\mathbf{E}}_{kk}).  
\end{equation}
The mutual information is obtained by
\begin{equation} \label{eq:mutualinformation5_lemma3}
    I(\boldsymbol{\varsigma}_{k};\mathbf{y}_{k}, \bar{\mathbf{E}}_{kk}) = h(\boldsymbol{\varsigma}_{k}) -h(\boldsymbol{\varsigma}_{k}|\mathbf{y}_{k}, \bar{\mathbf{E}}_{kk}). 
\end{equation}
If we suboptimally choose $\boldsymbol{\varsigma}_{k}$ as Gaussian, where $\boldsymbol{\varsigma}_{k} \sim \mathcal{N}_{\mathbb{C}}(0,\mathbf{I}_{M})$, then the left term is equal to
\begin{equation} \label{eq:differentialentropyleft5_lemma3}
    h(\boldsymbol{\varsigma}_{k}) = \log_{2} | \pi e \mathbf{I}_{M} |.
\end{equation}
To bound the second term in \eqref{eq:mutualinformation5_lemma3}, suppose we compute the LMMSE estimate $\hat{\boldsymbol{\varsigma}}_{k}$ of $\boldsymbol{\varsigma}_{k}$ based on $\mathbf{y}_{k}$ as we have a non-Gaussian observation equation, which is $\boldsymbol{\varsigma}_{k};\mathbf{y}_{k}| \bar{\mathbf{E}}_{kk}$. Therefore, the second differential entropy term is bounded above by covariance of estimation error $\Tilde{\boldsymbol{\varsigma}}_{k}= \boldsymbol{\varsigma}_{k} - \hat{\boldsymbol{\varsigma}}_{k}$, 
\begin{align} \nonumber
    h(\boldsymbol{\varsigma}_{k}|\mathbf{y}_{k}, \bar{\mathbf{E}}_{kk})  &=  
    h(\boldsymbol{\varsigma}_{k}-\hat{\boldsymbol{\varsigma}}_{k}|\mathbf{y}_{k}, \bar{\mathbf{E}}_{kk}) 
    \leq h(\boldsymbol{\varsigma}_{k}-\hat{\boldsymbol{\varsigma}}_{k}) \\ &=    \log_{2} | \pi e (\mathbf{C}_{\Tilde{\boldsymbol{\varsigma}}_{k}}  )|. \label{eq:differentialentropyright5}
\end{align}
We can write $
    \mathbf{C}_{\Tilde{\boldsymbol{\varsigma}}_{k}} = \mathbf{C}_{\boldsymbol{\varsigma}_{k}} - \mathbf{C}_{\boldsymbol{\varsigma}_{k}\mathbf{y}_{k}}     \mathbf{C}_{\mathbf{y}_{k}\mathbf{y}_{k}}^{-1} \mathbf{C}_{\boldsymbol{\varsigma}_{k}\mathbf{y}_{k}}
=    \mathbf{C}_{\Tilde{\boldsymbol{\varsigma}}_{k}} = \mathbf{I}_{M} - \bar{\mathbf{E}}_{kk}^{\rm{H}} \mathbf{C}_{\mathbf{n}''_{k}}^{-1} \bar{\mathbf{E}}_{kk},
$
where \textcolor{black}{$\mathbf{C}_{\boldsymbol{\varsigma}_{k}},\mathbf{C}_{\boldsymbol{\varsigma}_{k}\mathbf{y}_{k}},\mathbf{C}_{\mathbf{y}_{k}\mathbf{y}_{k}}$ are the covariance of data, cross-covariance of data with the received signal and the covariance of the received signal respectively.} We also have $
     \mathbf{C}_{\mathbf{n}''_{k}} =  \mathbb{E}\left\{   \mathbf{n}''_{k} \mathbf{n}''^{\rm{H}}_{k} \right\}. 
$
Therefore, the second differential entropy term is bounded above by 
\begin{equation} \label{eq:differentialentropyright5_lemma3}
    h(\boldsymbol{\varsigma}_{k}|\mathbf{y}_{k}, \bar{\mathbf{E}}_{kk}) \leq    \log_{2} | \pi e (\mathbf{I}_{M} - \bar{\mathbf{E}}_{kk}^{\rm{H}} \mathbf{C}_{\mathbf{n}''_{k}}^{-1} \bar{\mathbf{E}}_{kk}  )|.
\end{equation}
Combining \eqref{eq:differentialentropyleft5_lemma3} and \eqref{eq:differentialentropyright5_lemma3} and using matrix inversion lemma, we obtain
\begin{equation}
    I(\boldsymbol{\varsigma}_{k};\mathbf{y}_{k} , \bar{\mathbf{E}}_{kk}) \geq    \log_{2} |  (\mathbf{I}_{M} + \bar{\mathbf{E}}_{kk}^{\rm{H}} \mathbf{C}_{\mathbf{n}''_{k}}^{-1} \bar{\mathbf{E}}_{kk}  )|.
\end{equation}
Using this relation in \eqref{capacity_definition_lemma3} and multiplying with the pre-log factor, we obtain \eqref{eq:lemma_main_result_general_combiner}.

\subsection{Derivation of the MMSE Precoder} \label{appendix: MMSE Precoder}

We will derive the MMSE precoder considering a virtual uplink and use a uplink-downlink duality similar to \cite[Ch. 6]{demir2021foundations}. As we want to send separate streams in the downlink, we consider transmitting separate streams to the APs in the uplink. We write the received signal at AP $l$ with precoders $ \boldsymbol{\Gamma}_{ji}$ as
\begin{align} \label{downlink_transmitted_signal_2_appendix}
      \mathbf{y}_{l} &=  \mathbf{H}_{lk}  \boldsymbol{\Gamma}_{lk} \boldsymbol{\varsigma}_{k} + \sum_{j=1}^{L}\sum_{i=1, i \neq k}^{K}   \mathbf{H}_{ji} \boldsymbol{\Gamma}_{ji} \boldsymbol{\varsigma}_{i} +\mathbf{n}_{l}.
\end{align}
To detect the selected part of the stream $\boldsymbol{\Gamma}_{lk} \boldsymbol{\varsigma}_{k}$, AP $l$ applies a combiner $\mathbf{U}_{lk}$ and the filtered received signal becomes
\begin{align}\label{downlink_transmitted_signal_2_after_combiner}
     \mathbf{U}_{lk}^{\mathrm{H}}\mathbf{y}_{l} &=  \mathbf{U}_{lk}^{\mathrm{H}}\mathbf{H}_{lk}  \boldsymbol{\Gamma}_{lk} \boldsymbol{\varsigma}_{k} + \sum_{j=1}^{L}\sum_{i=1, i \neq k}^{K}  \mathbf{U}_{lk}^{\mathrm{H}} \mathbf{H}_{ji} \boldsymbol{\Gamma}_{ji} \boldsymbol{\varsigma}_{i} +\mathbf{U}_{lk}^{\mathrm{H}}\mathbf{n}_{l}.
\end{align}
We can write the MSE in the signal reception at AP $l$ as
\begin{align}    
    \mathrm{MSE_{lk}} = \mathbb{E}\left\{ | \mathbf{U}_{lk}^{\mathrm{H}}\mathbf{y}_{l}  - \boldsymbol{\varsigma}_{lk}|^{2} \Big| \Omega\right\}, 
\end{align}
where $\Omega$ denotes the side information available at AP $l$ and $\boldsymbol{\varsigma}_{lk}=\boldsymbol{\Gamma}_{lk}\boldsymbol{\varsigma}_{k}$. For CSI-sharing, we have $\Omega =\{ \hat{\mathbf{H}}_{i}$ for $i=1,\hdots,K\}$ and for no CSI-sharing we have $\Omega =\{ \hat{\mathbf{H}}_{li}$ for $ i=1,\hdots,K\}$ for AP l which result in \eqref{eq: MMSE Precoder for Option 3} and \eqref{eq: MMSE Precoder for Option 2} respectively.
We first open this relation as 
\begin{align}    
    &\mathrm{MSE_{lk}} = \mathbf{U}_{lk}^{\mathrm{H}} \mathbf{C}_{\mathbf{y}_{l}\mathbf{y}_{l}|\Omega} \mathbf{U}_{lk} - \mathbf{U}_{lk}^{\mathrm{H}} \mathbf{C}_{\mathbf{y}_{l}\boldsymbol{\varsigma}_{lk}|\Omega} -  \mathbf{C}_{\boldsymbol{\varsigma}_{lk}\mathbf{y}_{l}|\Omega} \mathbf{U}_{lk} + \mathbf{I} \nonumber \\
    &= \mathbf{I} - \mathbf{C}_{\boldsymbol{\varsigma}_{lk}\mathbf{y}_{l}|\Omega} \mathbf{C}_{\mathbf{y}_{l}\mathbf{y}_{l}|\Omega}^{-1} \mathbf{C}_{\mathbf{y}_{l}\boldsymbol{\varsigma}_{lk}|\Omega} \nonumber \\ &+ [(\mathbf{U}_{lk} -\mathbf{C}_{\mathbf{y}_{l}\mathbf{y}_{l}|\Omega}^{-1}\mathbf{C}_{\mathbf{y}_{l}\boldsymbol{\varsigma}_{lk}|\Omega})^{\mathrm{H}} \mathbf{C}_{\mathbf{y}_{l}\mathbf{y}_{l}|\Omega} (\mathbf{U}_{lk} -\mathbf{C}_{\mathbf{y}_{l}\mathbf{y}_{l}|\Omega}^{-1}\mathbf{C}_{\mathbf{y}_{l}\boldsymbol{\varsigma}_{lk}|\Omega})],
\end{align}
Here, the covariances are defined as $\mathbf{C}_{\mathbf{a}\mathbf{b}|\Omega}= \mathbb{E}\left\{\mathbf{a}\mathbf{b}^{\mathrm{H}}|\Omega\right\}$ for any $\mathbf{a}$ and $\mathbf{b}$ vectors.
The expectations are taken with the given side information for each case. As the last term is non-negative, the linear filter that minimizes the MSE is 
\begin{align}    \mathbf{U}_{lk}=\mathbf{C}_{\mathbf{y}_{l}\mathbf{y}_{l}|\Omega}^{-1}\mathbf{C}_{\mathbf{y}_{l}\boldsymbol{\varsigma}_{lk}|\Omega}.
\end{align}

By using the uplink-downlink duality, we can conclude that this relation can be used for the MMSE precoder too. Therefore, the precoder can be obtained as given in \eqref{eq: MMSE Precoder for Option 3}. Changing the available CSI at the APs changes the result of the precoders. If there is no CSI sharing between the APs, then the precoder will result as in \eqref{eq: MMSE Precoder for Option 2}. 

\bibliographystyle{IEEEtran}
\bibliography{cellfree}

@ARTICLE{Hassibi2003a,
  author={Hassibi, B. and Hochwald, B.M.},
  journal={IEEE Trans. Inf. Theory}, 
  title={How much training is needed in multiple-antenna wireless links?}, 
  year={2003},
  volume={49},
  number={4},
  pages={951-963}}

@article{mai2020downlink,
  title={Downlink spectral efficiency of cell-free massive {MIMO} systems with multi-antenna users},
  author={Mai, T. C. and Ngo, H. Q. and Duong, T. Q.},
  journal={IEEE Trans. Commun.},
  volume={68},
  number={8},
  pages={4803--4815},
  year={2020},
  publisher={IEEE}
}

@inproceedings{li2016massive,
  title={Massive {MIMO} with multi-antenna users: When are additional user antennas beneficial?},
  author={Li, X. and Bj{\"o}rnson, E. and Zhou, S. and Wang, J.},
  booktitle={Proc. 23rd Int. Conf. on Telecommun. (ICT)},
  pages={1--6},
  year={2016},
  organization={}
}

@article{wang2022uplink,
  title={Uplink performance of cell-free massive {MIMO} with multi-antenna users over jointly-correlated Rayleigh fading channels},
  author={Wang, Z. and Zhang, J. and Ai, B. and Yuen, C. and Debbah, M.},
  journal={IEEE Trans. Wireless Commun.},
  volume={21},
  number={9},
  pages={7391--7406},
  year={2022},
  publisher={IEEE}
}

@article{ngo2017no,
  title={No downlink pilots are needed in {TDD} massive {MIMO}},
  author={Ngo, H. Q. and Larsson, E. G.},
  journal={IEEE Trans. Wireless Commun.},
  volume={16},
  number={5},
  pages={2921--2935},
  year={2017},
  publisher={IEEE}
}

@article{demir2021foundations,
  title={Foundations of user-centric cell-free massive {MIMO}},
  author={{\"O}. T. Demir and E. Bj{\"o}rnson and L. Sanguinetti},
  journal={Foundations and Trends{\textregistered} in Signal Processing},
  volume={14},
  number={3-4},
  pages={162--472},
  year={2021},
  publisher={Now Publishers, Inc.}
}

@book{kay1993fundamentals,
  title={Fundamentals of statistical signal processing: estimation theory},
  author={Kay, S. M.},
  year={1993},
  publisher={Prentice-Hall, Inc.}
}

@article{bjornson2017massive,
  title={Massive {MIMO} networks: Spectral, energy, and hardware efficiency},
  author={Bj{\"o}rnson, E. and Hoydis, J. and Sanguinetti, L.},
  journal={Foundations and Trends{\textregistered} in Signal Processing},
  volume={11},
  number={3-4},
  pages={154--655},
  year={2017},
  publisher={Now Publishers, Inc.}
}

@article{buzzi2017cell,
  title={Cell-free massive {MIMO}: User-centric approach},
  author={S. Buzzi and C. D’Andrea},
  journal={IEEE Wireless Commun. Lett.},
  volume={6},
  number={6},
  pages={706--709},
  year={2017},
  publisher={IEEE}
}

@article{buzzi2019user,
  title={User-centric 5{G} cellular networks: Resource allocation and comparison with the cell-free massive {MIMO} approach},
  author={Buzzi, S. and D’Andrea, C. and Zappone, A. and D’Elia, C.},
  journal={IEEE Trans. Wireless Commun.},
  volume={19},
  number={2},
  pages={1250--1264},
  year={2019},
  publisher={IEEE}
}

@inproceedings{mai2019uplink,
  title={Uplink spectral efficiency of cell-free massive {MIMO} with multi-antenna users},
  author={Mai, T. C. and Ngo, H. Q. and Duong, T. Q.},
  booktitle={Proc. IEEE Int. Conf. Recent Adv. Signal Process. Telecommun. Comput. (SigTelCom)},
  pages={126--129},
  year={2019},
  organization={}
}

@article{zhang2019performance,
  title={On the performance of cell-free massive {MIMO} with low-resolution {ADC}s},
  author={Zhang, Y. and Zhou, M. and Qiao, X. and Cao, H. and Yang, L.},
  journal={IEEE Access},
  volume={7},
  pages={117968--117977},
  year={2019},
  publisher={IEEE}
}

@article{zhou2021sum,
  title={Sum-{SE} for multigroup multicast cell-free massive {MIMO} with multi-antenna users and low-resolution {DAC}s},
  author={Zhou, M. and Yang, L. and Zhu, H.},
  journal={IEEE Wireless Commun. Lett.},
  volume={10},
  number={8},
  pages={1702--1706},
  year={2021},
  publisher={IEEE}
}

@ARTICLE{sun2023uplink,
  author={Sun, Q. and Ji, X. and Wang, Z. and Chen, X. and Yang, Y. and Zhang, J. and Wong, K.},
  journal={IEEE Trans. Commun.}, 
  title={Uplink Performance of Hardware-Impaired Cell-Free Massive {MIMO} with Multi-Antenna Users and Superimposed Pilots}, 
  year={2023},
  volume={},
  number={},
  pages={1-1},
  doi={10.1109/TCOMM.2023.3301071}}

@ARTICLE{9650567,
  author={H. A. Ammar and R. Adve and S. Shahbazpanahi and G. Boudreau and K. V. Srinivas},
  journal={IEEE Commun. Surv. Tutor.}, 
  title={User-Centric Cell-Free Massive {MIMO} Networks: A Survey of Opportunities, Challenges and Solutions}, 
  year={2022},
  volume={24},
  number={1},
  pages={611-652},
  doi={10.1109/COMST.2021.3135119}}

@article{chen2022survey,
  title={A survey on user-centric cell-free massive {MIMO} systems},
  author={S. Chen and J. Zhang and J. Zhang and E. Bj{\"o}rnson and B. Ai},
  journal={Dig. Commun. Netw.},
  volume={8},
  number={5},
  pages={695--719},
  year={2022},
  publisher={Elsevier}
}

@article{interdonato2019ubiquitous,
  title={Ubiquitous cell-free massive {MIMO} communications},
  author={G. Interdonato and E. Bj{\"o}rnson and H. Q. Ngo and P. Frenger and E. G. Larsson},
  journal={EURASIP J. Wirel. Commun. Netw.},
  number={1},
  pages={1--13},
  year={2019},
  publisher={Springer}
}

@article{wu2017novel,
  title={Novel insight into multi-user channels with multi-antenna users},
  author={Wu, Xiaoyong and Liu, Danpu},
  journal={IEEE Commun. Lett.},
  volume={21},
  number={9},
  pages={1961--1964},
  year={2017},
  publisher={IEEE}
}

@ARTICLE{Interdonato2021a,
  author={Interdonato, Giovanni and Ngo, Hien Quoc and Larsson, Erik G.},
  journal={IEEE Trans. Commun.}, 
  title={Enhanced Normalized Conjugate Beamforming for Cell-Free Massive {MIMO}}, 
  year={2021},
  volume={69},
  number={5},
  pages={2863-2877}}

@ARTICLE{Chen2018a,
  author={Chen, Zheng and Björnson, Emil},
  journal={IEEE Trans. Commun.}, 
  title={Channel Hardening and Favorable Propagation in Cell-Free Massive {MIMO} With Stochastic Geometry}, 
  year={2018},
  volume={66},
  number={11},
  pages={5205-5219}}

@inproceedings{dovelos2020massive,
  title={Massive {MIMO} with multi-antenna users under jointly correlated Ricean fading},
  author={Dovelos, Konstantinos and Matthaiou, Michail and Ngo, Hien Quoc and Bellalta, Boris},
  booktitle={Proc. IEEE Int. Conf. Commun. (ICC)},
  pages={1--6},
  year={2020},
  organization={}
}

@article{sutton2021hardening,
  title={Hardening the channels by precoder design in massive {MIMO} with multiple-antenna users},
  author={Sutton, James AC and Ngo, Hien Quoc and Matthaiou, Michail},
  journal={IEEE Trans. Veh. Technol.},
  volume={70},
  number={5},
  pages={4541--4556},
  year={2021},
  publisher={IEEE}
}

@article{wang2020uplink,
  title={Uplink performance of cell-free massive {MIMO} over spatially correlated Rician fading channels},
  author={Wang, Zhe and Zhang, Jiayi and Bj{\"o}rnson, Emil and Ai, Bo},
  journal={IEEE Commun. Lett.},
  volume={25},
  number={4},
  pages={1348--1352},
  year={2020},
  publisher={IEEE}
}

@inproceedings{stankovic2004multi,
  title={Multi-user {MIMO} downlink precoding for users with multiple antennas},
  author={Stankovic, Veljko and Haardt, Martin},
  booktitle={Proc. of the 12-th Meeting of the Wireless World Research Forum (WWRF), Toronto, ON, Canada},
  volume={10},
  pages={12--14},
  year={2004}
}

@inproceedings{bandemer2006linear,
  title={Linear MMSE multi-user {MIMO} downlink precoding for users with multiple antennas},
  author={Bandemer, Bernd and Haardt, Martin and Visuri, Samuli},
  booktitle={Proc. IEEE 17th Int. Symp. Pers., Indoor Mobile Radio Commun.},
  pages={1--5},
  year={2006},
  organization={}
}

@article{spencer2004zero,
  title={Zero-forcing methods for downlink spatial multiplexing in multiuser {MIMO} channels},
  author={Spencer, Quentin H and Swindlehurst, A Lee and Haardt, Martin},
  journal={IEEE Trans. signal process.},
  volume={52},
  number={2},
  pages={461--471},
  year={2004},
  publisher={IEEE}
}

@article{ganesan2024cell,
  title={Cell-Free Massive {MIMO} with Multi-Antenna Users and Phase Misalignments: A Novel Partially Coherent Transmission Framework},
  author={Ganesan, Unnikrishnan Kunnath and Vu, Tung Thanh and Larsson, Erik G},
  journal={arXiv preprint arXiv:2403.00674},
  year={2024}
}

@ARTICLE{8799031,
  author={Interdonato, Giovanni and Ngo, Hien Quoc and Frenger, Pål and Larsson, Erik G.},
  journal={IEEE Trans. Wireless Commun.}, 
  title={Downlink Training in Cell-Free Massive {MIMO}: A Blessing in Disguise}, 
  year={2019},
  volume={18},
  number={11},
  pages={5153-5169},
  doi={10.1109/TWC.2019.2933831}}

@misc{kama2024downlinkpilotsessentialcellfree,
      title={Downlink Pilots are Essential for Cell-Free Massive {MIMO} with Multi-Antenna Users}, 
      author={Eren Berk Kama and Junbeom Kim and Emil Björnson},
      year={2024},
      eprint={2404.18516},
      archivePrefix={arXiv},
      primaryClass={eess.SP},
      url={https://arxiv.org/abs/2404.18516}, 
}

@INPROCEEDINGS{Kanno2022Fronthaul,
  author={Kanno, Issei and Ito, Masaaki and Ohseki, Takeo and Yamazaki, Kosuke and Kishi, Yoji and Choi, Thomas and Chen, Wei-Yu and Molisch, Andreas F.},
  booktitle={2022 IEEE 95th Vehicular Technology Conference}, 
  title={Fronthaul Load-Reduced Scalable Cell-Free massive {MIMO} by Uplink Hybrid Signal Processing}, 
  year={2022},
  volume={},
  number={},
  pages={1-5},
  doi={10.1109/VTC2022-Spring54318.2022.9860485}}

@article{shaik2025over,
  title={Over-the-Air Fronthaul Signaling for Uplink Cell-Free Massive {MIMO} Systems},
  author={Shaik, Zakir Hussain and Thoota, Sai Subramanyam and Bj, Emil and Larsson, Erik G and others},
  journal={IEEE Trans. Wireless Commun.},
  year={2025},
  publisher={IEEE}
}

@INPROCEEDINGS{10571171,
  author={Qian, Jun and Zhang, Chi and Letaief, Khaled B. and Murch, Ross},
  booktitle={2024 IEEE Wireless Communications and Networking Conference (WCNC)}, 
  title={The Effect of Spatial Correlation and Mutual Coupling on Cell-Free Massive {MIMO}}, 
  year={2024},
  volume={},
  number={},
  pages={01-06},
  doi={10.1109/WCNC57260.2024.10571171}}

@ARTICLE{9160993,
  author={Shen, Kaiming and Cheng, Hei Victor and Chen, Xihan and Eldar, Yonina C. and Yu, Wei},
  journal={IEEE Trans. Commun.}, 
  title={Enhanced Channel Estimation in Massive MIMO via Coordinated Pilot Design}, 
  year={2020},
  volume={68},
  number={11},
  pages={6872-6885},
  doi={10.1109/TCOMM.2020.3014680}}

@ARTICLE{6415397,
  author={Yin, Haifan and Gesbert, David and Filippou, Miltiades and Liu, Yingzhuang},
  journal={IEEE J. Sel. Areas Commun.}, 
  title={A Coordinated Approach to Channel Estimation in Large-Scale Multiple-Antenna Systems}, 
  year={2013},
  volume={31},
  number={2},
  pages={264-273},
  doi={10.1109/JSAC.2013.130214}}

@ARTICLE{9110802,
  author={Liu, Heng and Zhang, Jiayi and Jin, Shi and Ai, Bo},
  journal={IEEE Trans. Veh. Technol.}, 
  title={Graph Coloring Based Pilot Assignment for Cell-Free Massive MIMO Systems}, 
  year={2020},
  volume={69},
  number={8},
  pages={9180-9184},
  doi={10.1109/TVT.2020.3000496}}

@ARTICLE{11044434,
  author={Nassirpour, Sajjad and Nguyen, Toan-Van and Ngo, Hien Quoc and Tran, Le-Nam and Ratnarajah, Tharmalingam and Nguyen, Duy H. N.},
  journal={IEEE J. Sel. Topics Signal Process.}, 
  title={Variational Bayesian Channel Estimation and Data Detection for Cell-Free Massive MIMO with Low-Resolution Quantized Fronthaul Links}, 
  year={2025},
  volume={},
  number={},
  pages={1-15},
  doi={10.1109/JSTSP.2025.3579644}}

@ARTICLE{8094949,
  author={Björnson, Emil and Hoydis, Jakob and Sanguinetti, Luca},
  journal={IEEE Transactions on Wireless Communications}, 
  title={Massive MIMO Has Unlimited Capacity}, 
  year={2018},
  volume={17},
  number={1},
  pages={574-590},
  doi={10.1109/TWC.2017.2768423}}

\begin{IEEEbiography}[{\includegraphics[width=1in,height=1.25in,clip,keepaspectratio]{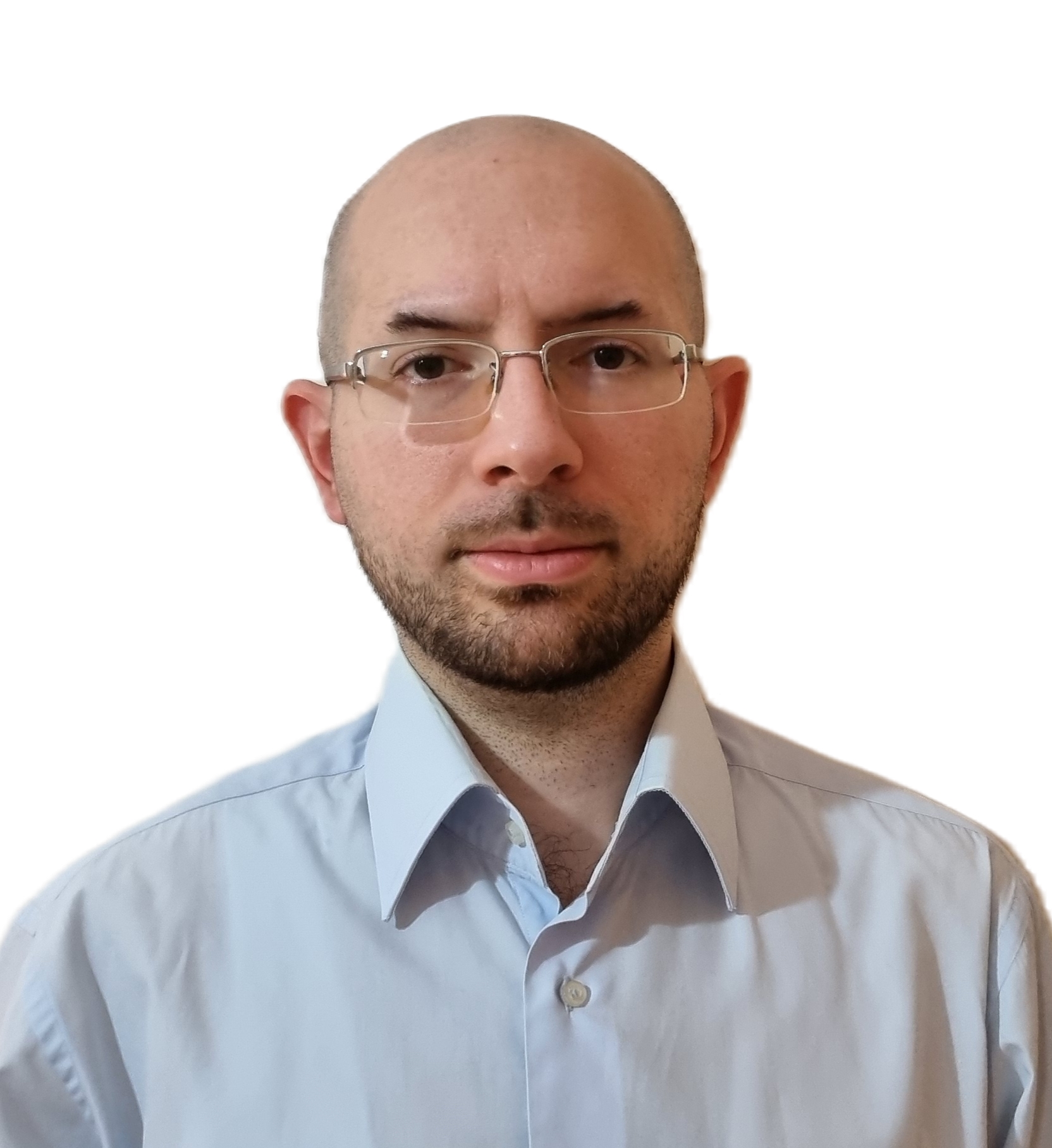}}]{Eren Berk Kama (Member, IEEE)} received the B.S. and MSc. degrees in electrical and electronics engineering from Middle East Technical University, Ankara, Turkey, in 2019 and 2022, respectively. Since 2022, he has been a Ph.D. student at KTH Royal Institute of Technology, Stockholm, Sweden. His research interests are wireless communications and signal processing.
\end{IEEEbiography}

\begin{IEEEbiography}[{\includegraphics[width=1in,height=1.25in,clip,keepaspectratio]{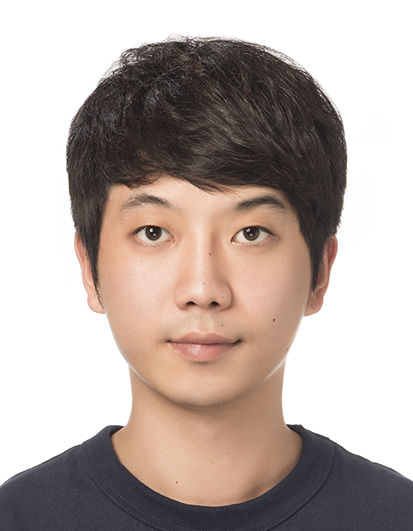}}]{Junbeom Kim (Member, IEEE)} received the B.S. and Ph.D. degrees in electronics and information engineering from Jeonbuk National University, Jeonju, Korea, in 2018 and 2023, respectively. In 2023, he was a Post-Doctoral Research Fellow at KTH Royal Institute of Technology, Stockholm, Sweden. Since 2023, he has been with Gyeongsang National University, Jinju, Korea, where he is currently an Assistant Professor with the Department of AI Information Engineering. His research interests include wireless communication technologies in PHY layer, signal processing optimization and machine learning.
\end{IEEEbiography}

\begin{IEEEbiography}[{\includegraphics[width=1in,height=1.25in,clip,keepaspectratio]{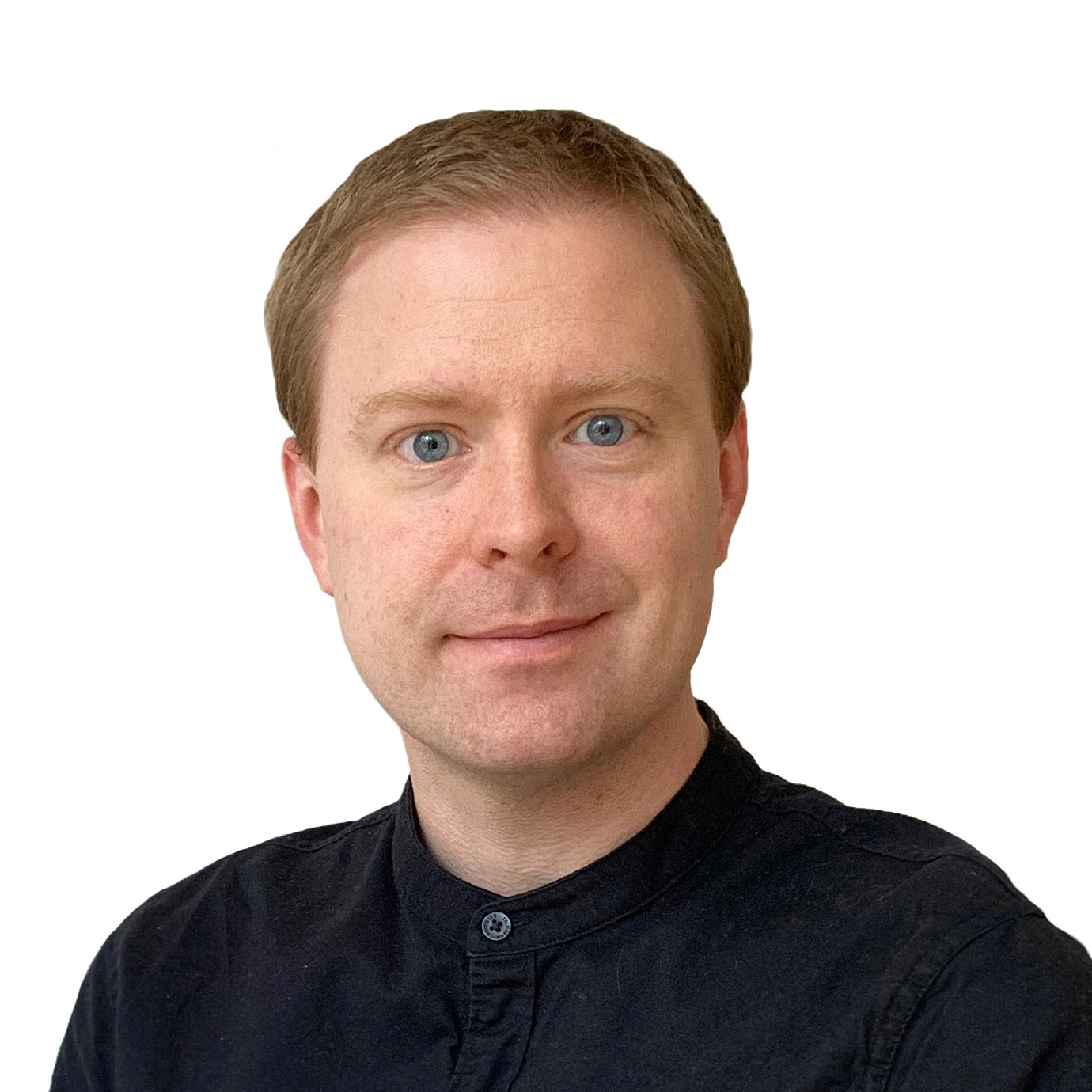}}]{Emil Bj{\"o}rnson (Fellow, IEEE)}  received the M.S. degree in engineering mathematics from Lund University, Sweden, in 2007, and the Ph.D. degree in telecommunications from the KTH Royal Institute of Technology, Sweden, in 2011.
 He is now a Professor of Wireless Communication at the KTH Royal Institute of Technology, Stockholm, Sweden. He is an IEEE Fellow, Digital Futures Fellow, Wallenberg Academy Fellow, and Clarivate Highly Cited Researcher. He has a podcast and YouTube channel called Wireless Future. His research focuses on multi-antenna communications and radio resource management, using methods from communication theory, signal processing, and machine learning. He has authored four textbooks and published much simulation code.
 
He has received the 2018 and 2022 IEEE Marconi Prize Paper Awards in Wireless Communications, the 2019 EURASIP Early Career Award, the 2019 IEEE Communications Society Fred W. Ellersick Prize, the 2019 IEEE Signal Processing Magazine Best Column Award, the 2020 Pierre-Simon Laplace Early Career Technical Achievement Award, the 2020 CTTC Early Achievement Award, the 2021 IEEE ComSoc RCC Early Achievement Award, the 2023 IEEE ComSoc Outstanding Paper Award, and the 2024 IEEE Stephen O. Rice Prize.
 \end{IEEEbiography}

\end{document}